\documentclass[useAMS,usenatbib]{mn2e}

\usepackage[pdftex]{color}
\usepackage[pdftex]{graphicx}
\usepackage[draft]{hyperref}
\usepackage[fleqn]{amsmath}     
\usepackage{amssymb}

\newcommand{\kms}{km\,s$^{-1}$}

\begin{document}

\title[A discrete chemo-dynamical model of NGC~5846]{A discrete chemo-dynamical model of the giant elliptical galaxy NGC~5846: dark matter fraction, internal rotation and velocity anisotropy out to six effective radii} 
\author[Zhu., et al]{%
	Ling Zhu$^1$\thanks{E-mail: lzhu@mpia.de}, Aaron~J.~Romanowsky$^{2,3}$, 
    	Glenn~van~de~Ven$^1$, R. J.~ Long$^{4,5}$, \and Laura L. Watkins$^6$, Vincenzo Pota$^7$, Nicola R. Napolitano$^7$, Duncan~A.~Forbes$^8$,\and Jean Brodie$^9$, Caroline Foster$^{10}$,   \\
    	$^1$Max Planck Institute for Astronomy, K\"onigstuhl 17, 69117 Heidelberg, Germany \\
	$^2$Department of Physics \& Astronomy, San Jos\'e State University, One Washington Square, San Jose, CA 95192, USA \\
    	$^3$University of California Observatories, 1156 High Street, Santa Cruz, CA 95064, USA\\
    	$^4$National Astronomical Observatories, Chinese Academy of Sciences, A 20 Datun Rd, Chaoyang District, Beijing 100012, China\\
        $^5$Jodrell Bank Centre for Astrophysics, School of Physics and Astronomy, The University of Manchester, Oxford Road, Manchester M13 9PL, UK\\
        $^6$Space Telescope Science Institute, 3700 San Martin Drive, Baltimore, MD 21218, USA \\
        $^7$INAF - Osservatorio Astronomico di Capodimonte, Salita Moiariello, 16, I-80131 Napoli, Italy\\
        $^8$Centre for Astrophysics \& Supercomputing, Swinburne University, Hawthorn, VIC 3122, Australia \\
        $^9$University of California Observatories, 1156 High Street, Santa Cruz, CA 95064, USA \\
        $^{10}$Australian Astronomical Observatory, PO Box 915, North Ryde, NSW 1670, Australia \\
	}

\maketitle 
\date{Accepted 0000 Month 00. Received 0000 Month 00; in original 0000 Month 00}

\pagerange{\pageref{firstpage}--\pageref{lastpage}} \pubyear{2014}

\label{firstpage}
\begin{abstract}
\begin{itemize}
We construct a suite of discrete chemo-dynamical models of the giant elliptical galaxy NGC 5846. These models are
a powerful tool to constrain both the mass distribution and internal dynamics of multiple tracer populations.
We use Jeans models to simultaneously fit stellar kinematics within the effective radius $R_{\rm e}$, planetary nebula (PN) radial velocities out to $3\, R_{\rm e}$, and globular cluster (GC) radial velocities and colours out to $6\,R_{\rm e}$. 
The best-fitting model is a cored DM halo which contributes $\sim 10\%$ of the total mass within $1\,R_{\rm e}$, and 
$67\% \pm 10\%$ within $6\,R_{\rm e}$, although a cusped DM halo is also acceptable.
The red GCs exhibit mild rotation with $v_{\rm max}/\sigma_0 \sim 0.3$ in the region $R > \,R_{\rm e}$, aligned with but counter-rotating to the stars in the inner parts, while the blue GCs and PNe kinematics are consistent with no rotation. The red GCs are tangentially anisotropic, the blue GCs are mildly radially anisotropic, and the PNe vary from radially to tangentially anisotropic from the inner to the outer region. This is confirmed by general made-to-measure models. 
The tangential anisotropy of the red GCs in the inner regions could stem from the
preferential destruction of red GCs on more radial orbits, while
their outer tangential anisotropy -- similar to the PNe in this region -- has no good explanation.
The mild radial anisotropy of the blue GCs is consistent with an accretion scenario.
\end{itemize}
\end{abstract}


\section{Introduction}
\label{S:intro}
Globular clusters (GCs) and planetary nebulae (PNe) are two powerful tracers of the dark matter distributions in giant elliptical galaxies.  At the same time, they are important tracers of the outer halo as a fossil record of a galaxy's formation history. 

How the outer haloes of giant ellipticals are assembled is still under debate.  
Numerical simulations of these galaxies have been carried out in a cosmological context \citep{Cooper2013,Wu2014,Rottgers2014,Pillepich2014}, and it has been found that the stellar mass surface density profiles show an upward break at the radius where stars accreted from previously distinct galaxies ({\it ex situ}) start to dominate over the stars formed in the galaxy itself ({\it in situ}). The break disappears in the most massive galaxies, where {\it ex situ} stars dominate at all radii.
The motions of {\it ex situ} stars are typically radially biased, while {\it in situ} stars can become tangentially biased if dissipation was significant during the later stages of assembly of the galaxy. Thus, in simulations, more massive galaxies with a large fraction of {\it ex situ} stars have a radially anisotropic velocity distribution outside of one effective radius. Tangential anisotropy is seen only for galaxies with high fractions of {\it in situ} stars, and is particularly rare in ``slow rotator'' ellipticals \citep{Wu2014}.
 
When comparing galaxies, or comparing observations and simulations, different tracers are often used and their kinematics are not necessarily the same.  
Elliptical galaxies usually have more than one population of GCs--metal rich (red) and metal poor (blue) (e.g. \citealt{Brodie2006}; \citealt{Brodie2012})--that exhibit different properties. The red GCs usually have a density distribution and velocity dispersion profile that follows the galaxy's field stars, while the blue GCs extend further spatially and have a higher velocity dispersion. The velocity anisotropy of the blue GCs is also different from the red GCs and stellar tracers \citep{Zhang2015, Pota2015b}. These different properties of the red and blue GCs indicate different formation scenarios. The expectation is that the red GCs are predominantly born {\it in situ}. The blue GCs are thought to be predominantly accreted from low mass galaxies. This accretion is evidenced by the known correlations between the average metallicity of a GC population and its host galaxy's mass \citep{Peng2006}. 
However, after the dissolution of some GCs by tidal forces, even the red GCs may obtain kinematic properties different from the galaxy's main stellar population. 
With increasing data volumes, recent analyses have examined the kinematics of each population of GCs separately
(e.g. \citealt{Schuberth2010,Pota2013, Pota2015a, Zhang2015}).

Modelling the kinematics of all tracers simultaneously, with distinct dynamical properties for each population, is a powerful technique for addressing degeneracies: tracers with different velocity anisotropies may help to alleviate the mass--anisotropy degeneracy, while kinematic constraints on different spatial scales can alleviate the luminous and dark matter mass degeneracy \citep{Napolitano2014,Agnello2014,Pota2015b}. 
However, several key aspects of this multi-population approach could be improved.
Models are typically restricted to spherical symmetry, and their fitting processes do not function with unbinned velocities and, so, do not maximize the constraining power of the data 
(see \citealt{Romanowsky2001,Bergond2006,Wu2006} for single-population examples).
GC subpopulations are separated through their distribution of colour or metallicity before dynamical modelling occurs.
In this context, the discrete chemo-dynamical modelling technique that we introduced in Zhu et al., 2016 (submitted)
is a powerful tool to separate multiple populations that overlap chemically, to investigate their dynamical properties and to constrain the underlying gravitational potential simultaneously. In this paper, we extend our axisymmetric Jeans modelling technique to include three dynamical populations (the stars/PNe and two GC subpopulations)
 and, in addition to the discrete data, also fit the integrated-light stellar kinematic data in the inner region. 
The only comparable work from the literature is the \citet{Oldham2016} spherical Jeans models of M87,
using stars, two GC subpopulations and satellite galaxies. 

The paper is organised as follows: in Section~\ref{S:obs} we present the observational data; in Section~\ref{method}, we describe our model algorithms; in Section~\ref{S:modelfits}, we present the results of modelling the kinematics; 
in Section~\ref{S:kin}, we present the internal rotation and velocity anisotropy profiles obtained by the chemo-dynamical Jeans models as well as those from the Made-to-Measure (M2M) models for each population. In Section~\ref{S:mass}, we show the mass profiles. 
We discuss the implications of the results in Section~\ref{S:discussion} and summarise in Section~\ref{S:conclusions}. In the Appendix, we show kinematic figures for different models and describe in more detail the M2M models we employed. 

\section{Data}
\label{S:obs}
NGC 5846 is a giant elliptical that is the brightest galaxy at the centre of a group at a distance of 24.2 Mpc away 
(\citealt{Tonry2001}, corrected by subtracting 0.06 mag from the distance modulus as suggested by \citealt{Mei2007}), 
so that $1'' = 117\,\mathrm{pc}$.  It appears nearly spherical in shape with Hubble type E0 and is kinematically
classified as a centrally-slow rotator \citep{Emsellem2011}.
The effective radius of the galaxy is $R_{\rm e} = 81'' \simeq 9.5 \, \mathrm{kpc}$. We adopt a major-axis position angle of $60^\circ$,
based on the stellar isophotes at large radii \citep{Kronawitter2000}.

Throughout, $(x, y, z)$ indicate the coordinates of the de-projected three-dimensional (3D) system; $r$ indicates the spherical shell radius. We place the projected major and minor axes of the galaxy along $(x', y')$, while $z'$ is along the line of sight; $R$ indicates the projected radius $R = \sqrt{x'^2 + y'^2}$.
We also use the projected semi-major elliptical annular radius defined as 
\begin{equation}
\label{eqn:RR}
R' \equiv {\rm sign}(x')  \times \sqrt{x'^2 + (y'/q)^2},
\end{equation} 
with an average flattening $q = 0.85$ 
(axis ratio between minor and major axis $b/a$)
adopted for all populations.

\subsection{Projected density profiles}
\label{SS:sd}
\begin{figure}
\centering\includegraphics[width=\hsize]{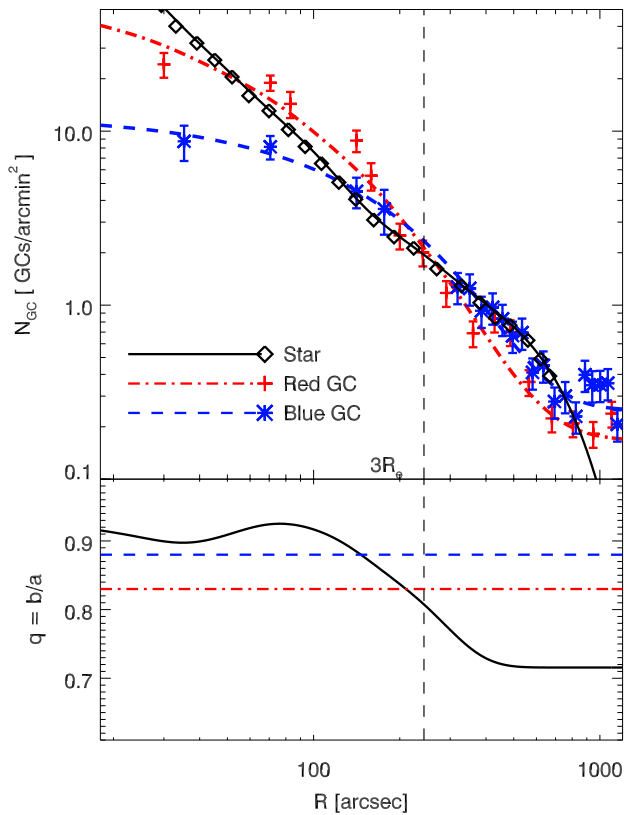}
\caption{Density profiles for NGC 5846. {\bf Top panel:} The stellar surface brightness and the GC surface number density profiles along the major axis. The black diamonds and solid curve are the observed data and MGE fit to the stellar surface brightness, respectively.
The red plus-symbols with error bars and dash-dotted curves are those for the red GCs, the blue asterisks with error bars and dashed curves are those for the blue GCs. 
The stellar data have been arbitrarily normalized for comparison with the GCs.
{\bf Bottom panel:} The axis ratio $q$ profiles. The black solid curve represents the stars. For the GCs, we adopt constant values of $q=0.83$ for the red GCs and 0.88 for the blue GCs, as indicated by the horizontal red dash-dotted and blue dashed lines. The vertical dashed line indicates the position of $3\,R_{\rm e}$ for the galaxy.} 
\label{fig:sdsb}
\end{figure}

Photometric observations provide the stellar surface brightness and the projected number density profiles for GCs, as shown in Fig.~\ref{fig:sdsb}. We take the observed $V$-band stellar surface brightness profile along the projected major axis of the galaxy from \citet{Kronawitter2000} (black diamonds), and we fit the surface brightness profile with a multi-Gaussian expansion (MGE; \citealt{Cappellari2002}; \citealt{Emsellem1994}). Eight Gaussian components are used for the best-fitting MGE to the surface brightness (black solid lines). This surface brightness profile has been normalized for comparison purposes. 
The surface number density profile of the PNe (not shown in the figure) is almost identical with the stellar surface brightness profile \citep{Coccato2009}. 

The surface number density profiles of red GCs (red pluses) and blue GCs (blue asterisks) are calculated from the GC photometric data. The photometric data were derived from
{\it Hubble Space Telescope} ({\it HST})/Wide-Field Planetary Camera 2 and Subaru/Suprime-Cam images,
as described in \citet{Napolitano2014}, but updated slightly here for ellipticity effects.
Briefly, the GC density profiles
 were computed for the {\it HST} and Subaru datasets separately and combined post-facto. 
The two datasets have different magnitude normalizations. Therefore, we first selected GCs 0.5 mag brighter than the turn-over-magnitude to eliminate this normalization difference. Then we divided the GCs into blue GCs and red GCs using $(g-i) < 0.90$ and $(g-i) > 0.95$ for the Subaru data sets, and using $(V - I ) < 1.05$ and $(V-I) > 1.09$ for the {\it HST} datasets. Note that the separation on $(g-i)$ is only used to determine the projected number density profiles to be used as model inputs. 

The surface number density profiles of red and blue GCs are described well by an MGE plus a constant background component, shown as red dash-dotted and blue dashed lines in Fig.~\ref{fig:sdsb}, respectively. The constant background term is 0.25 for the blue GCs and 0.167 for the red GCs, which will be subtracted from the profiles before the modeling. 

The projected density profiles are not spherical but spheroidal. 
In the bottom panel of Fig.~\ref{fig:sdsb}, we show the projected flattening profiles along the major axis for the three populations. The black solid line represents that of the stellar surface brightness, with values varying from
$q \sim 0.9$ in the center to $\sim 0.7$ in the outer regions. 
We adopt a constant flattening for the red and blue GCs: $q=0.83$ for red GCs (red dash-dotted line) and 0.88 for the blue GCs (blue dashed line), which are obtained from all the GC photometric data. 

The spectroscopic data we will use for dynamical constraints are far from complete and have complicated selection functions, leaving little ability to constrain the density or brightness profiles of the tracer populations. These profiles--determined beforehand from photometric data (not the spectroscopic data), and converted into MGE form--will provide input to our modelling process.

\subsection{Kinematic data}
\label{SS:kin}

\begin{figure*}
\centering\includegraphics[width=\hsize]{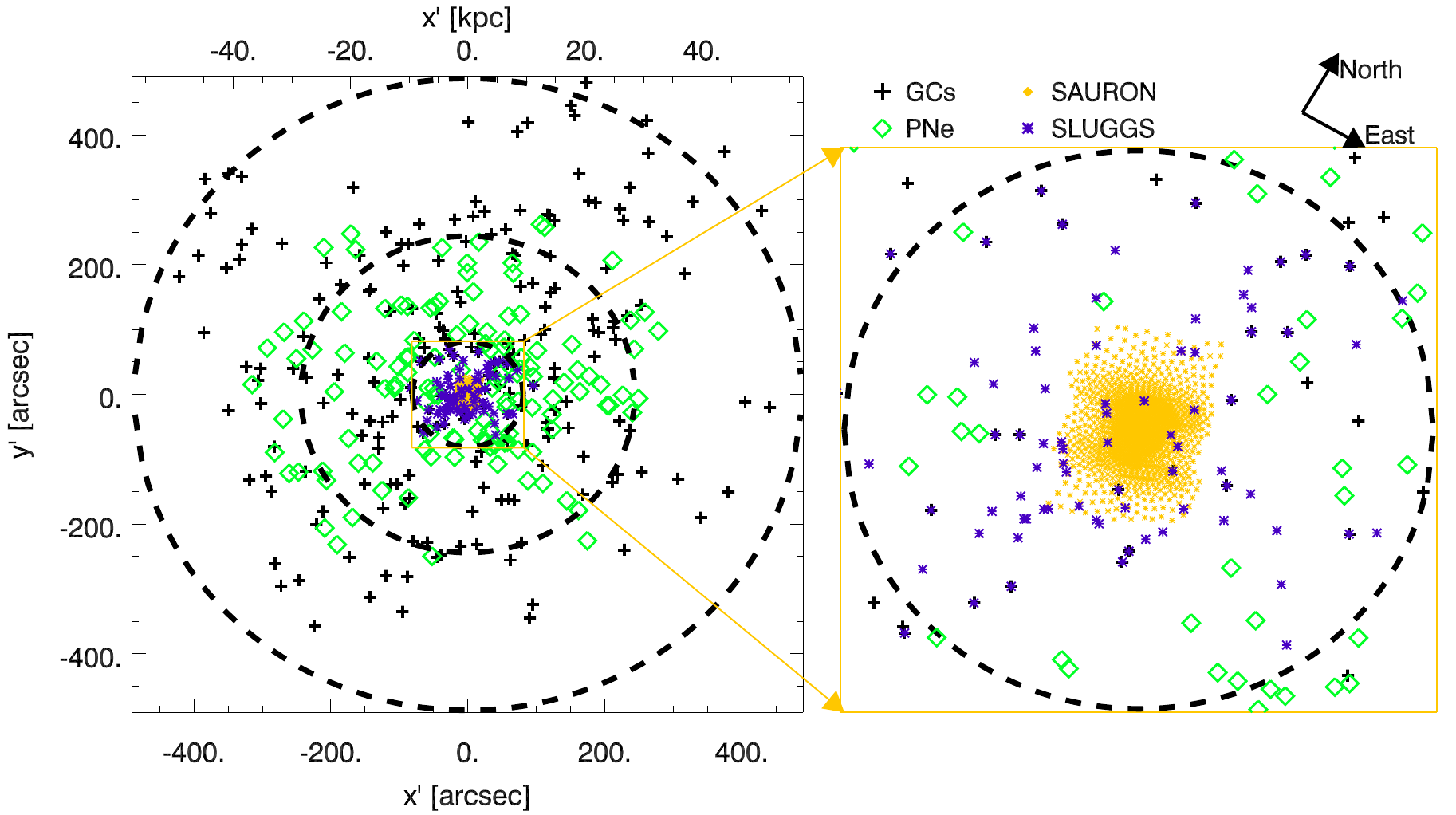}
\caption{The kinematic data in the plane of sky with $x'$ and $y'$ representing the photometric major and minor axes of the galaxy; the orientation of the figure is labeled in the top-right corner.  
The black pluses are the positions of GCs, the green diamonds are PNe, the purple asterisks are the integrated-light stellar kinematic observations from SLUGGS and the area filled with orange asterisks is the SAURON stellar data. The circles in the left panel indicate the scales of $1\,R_{\rm e}$, $3\,R_{\rm e}$ and $6\,R_{\rm e}$ for the galaxy. The right panel zooms in to the inner $1\,R_{\rm e}$ region. }
\label{fig:xyplot}
\end{figure*}

For NGC 5846, we have 
123 PNe with LOS velocity measurements extending to $\sim 37 \, \mathrm{kpc} \simeq 4\,R_{\rm e}$ from 
the Planetary Nebulae Spectrograph Survey \citep{Coccato2009}, and
214 GCs with line-of-sight (LOS) velocity measurements extending to $\sim 62\, \mathrm{kpc} \simeq 6.5\, R_{\rm e}$ from the SAGES Legacy Unifying Globulars and GalaxieS (SLUGGS) Survey \citep{Brodie2014}\footnote{\url{http://sluggs.ucolick.org}}.
The GCs are mostly the same data as described in \citet{Napolitano2014}, with the addition of $\sim$~20 new, preliminary velocities
from an observing run with Keck/DEIMOS on 2012-04-17
(to be discussed in full in D.\ Forbes et al., in preparation).

We also have integrated-light stellar kinematic measurements at 80 discrete positions extending to $\sim 1\, R_{\rm e}$, 
obtained by the SLUGGS survey. These measurements provide at each point the mean velocity and velocity dispersion which we will use for modelling, as well as the
 Gauss-Hermite coefficients $h_3$ and $h_4$. 
In our modelling, we discard the measurements that are around the satellite galaxy NGC~5846A toward the South.

The SAURON integral-field unit (IFU) data for NGC 5846 \citep{Emsellem2011}, extending to $\sim 0.5\,R_{\rm e}$, are also included in our model. 
The stellar velocity dispersions from the SAURON data are systematically higher than in SLUGGS by $\sim$~10~\kms\ --
a known issue found in comparing data from these two surveys (\citealt{Foster2016}; \citealt{Boardman2016}).
Here we add 10~\kms\ to the SLUGGS stellar velocity dispersions.
We check the effects on our models by alternatively subtracting 10~\kms\ from the SAURON data (see Section~\ref{S:mass}).

The kinematic data in the projected plane are shown in Fig.~\ref{fig:xyplot}. The black pluses are the positions of the GCs, the green diamonds are the PNe, the purple asterisks are the positions of stellar kinematics measurements from SLUGGS and the area filled with orange asterisks is the SAURON data.
The three dashed circles indicate scale radii of $1\,R_{\rm e}$, $3\,R_{\rm e}$ and $6\,R_{\rm e}$ for the galaxy. 

\section{Discrete chemo-dynamical models}
\label{method}

GCs themselves contribute very little to the total stellar luminosity of the galaxy, which
comes predominately from the stars in the central galactic structure. PNe simply mark dead low-mass stars, are thus assumed to follow the distribution of the main stellar population.
The stellar tracers (PNe + central galaxy stars), the red GCs and the blue GCs are three independent populations that trace the underlying galactic potential. 

We use the discrete chemo-dynamical modelling technique developed in Zhu et al., 2016 (submitted), 
and refer to this paper for the details of the technique. The dynamical properties of each population are described by an axisymmetric Jeans model with free velocity anisotropy and rotation parameters \citep{Cappellari2008}. The discrete data are modelled directly without spatial binning \citep{Watkins2013}, and the red and blue GCs are not separated via the colour distribution before the modelling. 
In this paper, we extend the method to also include integrated-light stellar kinematics. 

\subsection{Gravitational potential}
\label{SS:potential}
The gravitational potential is modelled by a combination of the stellar and dark matter mass distributions. We de-project the 2D MGE fit to the galaxy's surface brightness in $V$ band to get the 3D luminosity density of the galaxy. After assuming a constant stellar mass-to-light ratio $\Upsilon_V$, we obtain the 3D stellar mass density that generates the stellar contribution to the potential. 
When de-projecting the 2D MGE of the image to a 3D MGE for the luminosity, we assume that the line-of-sight inclination angle of the galaxy is $90^\circ$
(edge-on). The inclination angle is fixed in our model because, especially for a slowly rotating galaxy, it is poorly constrained by kinematic data \citep{vdB2009}. 

For the dark matter contribution, we adopt a generalized \citet{NFW1996} (NFW) density distribution
as in \citet{Zhao1996} 
\begin{equation}
	\rho(r) = \frac{\rho_{s} }{ (r/r_s)^{\gamma}(1 + (r/r_s)^{\eta})^{(3-\gamma)/\eta} },
	\label{eq:densgNFW}
\end{equation}
with $r^2 = x^2 + y^2 + z^2/p_z^2$ in the case of an oblate axisymmetric distribution. Since the flattening of the dark matter halo is, to a large degree, degenerate with its radial profile, line-of-sight data alone are expected to provide weak constraints if both the flattening and the radial profile of DM are left free. Numerical simulations show that dark matter halos are usually rounder than luminous matter \citep{Wu2014}, 
which, in the case of a massive elliptical galaxy, is not far from spherical, and hence
we adopt a spherical DM halo with $p_z = 1$. 

There are four free halo parameters: the scale radius $r_s$, the scale density $\rho_s$,  and the inner and outer density slopes $\gamma$ and $\eta$.
When ($\eta= 1,\, \gamma =1$), the halo reduces to a classic cusped profile \citep{NFW1996}, while, for ($\eta =2, \,\gamma =0$), there is a core in the centre.

We expand the density $\rho$ as an MGE to simplify various calculations such as the computation of the gravitational potential \citep{Emsellem1994} and the solution of the axisymmetric Jeans equation \citep{Cappellari2008}. 
The total density we use to generate the potential is just the combination of the 3D MGE of stellar and DM density.
The central black hole is ignored, a $\sim 10^8 \,M_{\odot}$ central black hole has an influence radius less than $1''$, which won't affect our results. 

\subsection{Tracer probabilities}
Consider a dataset of $N$ measured points such that the $i^{\rm th}$ point has sky coordinates $(x'_i, y'_i)$ and line-of-sight velocity $v_{z',i} \pm \delta{v_{z',i}}$ with measured metallicity $Z_i\pm\delta{Z_i}$.  

Multiple populations are permitted in the model, each with its own chemical, spatial and dynamical distributions. 
For each population $k$, we adopt a Gaussian metallicity distribution
(see \citealt{Forte2007} for an alternative distribution)
 with mean metallicity $Z_0^k$ and metallicity dispersion $\sigma_Z^{k} $ with both the mean and dispersion being taken as free parameters. 
 Here we do not use direct observations of metallicities (cf.\ \citealt{Usher2012}) but instead adopt $g-i$ colour as a proxy for metallicity.
 Note that the exact equivalence between colour and metallicity does not matter in this context;
 we are effectively modelling a two-component population of GC colours, which could in principle be re-interpreted according to some
 other stellar population parameters.

The metallicity probability of point $i$ in population $k$ is 
\begin{equation}
	\label{eq:pchm}
	P_{\mathrm{chm},i}^k = \frac{1}{\sqrt{2\pi[(\sigma_{Z}^k)^2+\delta{Z_i}^2]}}
		\exp\left[ -\frac12\frac{(Z_i-Z_0^k)^2}{[(\sigma_Z^k)^2 + \delta{Z_i}^2]}\right].
\end{equation}

Each population has its own spatial distribution through its observed projected number density $\Sigma^k(x',y')$, and so the spatial probability of point $i$ in population $k$ is 
\begin{equation}
	\label{eq:pspa}
	P_{\mathrm{spa},i}^k  = \frac{\Sigma^k(x'_i,y'_i)}{\Sigma_\mathrm{obj}(x'_i,y'_i)},
\end{equation}
where $\Sigma_\mathrm{obj} = \Sigma^1 + \Sigma^2 + \dots$ is the combined  density of all populations that belong to the object under consideration (for example, a galaxy).

The dynamical properties of each population are described by an axisymmetric Jeans model with a specific
density profile, velocity anisotropy and rotation parameter.
This model assumes that the velocity ellipsoid is aligned with the cylindrical coordinates and the velocity anisotropy in the meridional plane is constant. 
The dynamical probability of point $i$ in population $k$ for an assumed Gaussian velocity distribution is 
\begin{equation}
	\label{eq:pdyn}
	P_{\mathrm{dyn},i}^k  = \frac{1}{\sqrt{(\sigma^k_i)^2 + (\delta v_{z',i})^2}}
	\exp\left[ -\frac12\frac{(v_{z', i} - \mu^k_{i})^2 }{(\sigma^k_i)^2 + (\delta v_{z',i})^2}\right],
\end{equation}
where $\mu_i^k$ and $\sigma_i^k$ are the line-of-sight mean velocity and velocity dispersion as predicted by a dynamical model at the sky position $(x'_i,y'_i)$. To investigate the velocity anisotropy and the rotation properties, we include two free parameters for each population $k$ in an axisymmetric Jeans model (in a fixed potential): 
the constant velocity anisotropy in the meridional plane $\beta_z^{k}$ ($= 1 - \overline{v_z^2} / \overline{v_R^2}$) and the rotation parameter $\kappa^{k}$ ($= [\overline{v_{\phi}}] / ([\overline{v_{\phi}^2}] - [\overline{v_{R}^2}]  )^{1/2}$) \citep{Cappellari2008}.

In principle, we could let the density profiles of all populations be free within a model, i.e., the density profiles could be constrained by the actual spatial distribution of the tracers if we have sufficient tracers to sample adequately the true distributions. 
However, the kinematic data are usually far from complete, with complicated selection functions, and thus provide almost no constraints on the population density profiles. In practice, the projected density profiles are
taken from photometric data, with different populations being separated by metallicity (as shown in Fig.~\ref{fig:sdsb}) and used as model inputs. 

\subsubsection{Discrete GCs and PNe}
With discrete GC and PN data, the model can be considered as a generalization of the two-component discrete chemo-dynamical model (Zhu et al., 2016, submitted) to include three populations. 
However, the stellar tracers (PNe) and GCs are independent tracers from each other, hence the likelihood is
\begin{equation}
L_{i  \in {\rm GC}} = \sum_{k={\rm GC}} P_{\mathrm{chm},i}^k \,P_{\mathrm{spa},i}^k \, P_{\mathrm{dyn},i}^k 
\end{equation}
where point $i$ represents a GC, and the other subscripts indicate chemical, spatial or dynamical probability. 
Once the best-fitting parameters in a model have been obtained, the likelihood of a GC $i$ belonging to each GC population $k$ is
\begin{equation}
P_i^k = P_{{\rm spa},i}^k P_{{\rm chm},i}^k P_{{\rm dyn},i}^k,
\end{equation}
where $k$ can be either a red GC or a blue GC.
The relative probability $P_i^{'k}$, defined as
\begin{equation}
\label{eqn:Pik}
P_i^{'k} = P_i^k / \sum^{k={\rm GC}} P_i^k,
\end{equation}
can be used to identify a GC as being red or blue by utilising the end of run model. 

When the point $i$ is a PN, the likelihood is
\begin{equation}
L_{i \in {\rm PN}} =P_{\mathrm{dyn},i}^{\mathrm{star}},
\end{equation}
which is just the probability of an independent single component Jeans model. Both $L_{i  \in {\rm GC}}$ and $L_{i \in {\rm PN}}$ are evaluated in the same gravitational potential.

The total log likelihood of the discrete data is
\begin{multline}
\label{eqn:LogL}
\mathcal{L} = \sum_{i=1}^{N_{\rm GC}} \log{L_{i \in {\rm GC}}} +  \sum_{i=1}^{N_{\rm PN}} \log{L_{i \in {\rm PN}}}
\equiv \mathcal{L}_{\rm GC} + \mathcal{L}_{\rm PN}.
\end{multline}

\subsubsection{Integrated-light stellar kinematics}
\label{SS:combine}
The SAURON IFU data and the SLUGGS stellar kinematics provide the mean velocity and velocity dispersion  $ (\mu_j^{\rm s}\pm \delta \mu_{j}^{\rm s} , \sigma_j^{\rm s}\pm \delta \sigma_{j}^{\rm s}  )$ of the stars across the projected plane of the galaxy in bins with centroids ($x'_j, y'_j$).
 
We directly compare the mean velocity and velocity dispersion of the data with that predicted by the stellar tracer Jeans model, resulting in 
\begin{equation}
\chi^2_{\rm star} =  \sum_{j=1}^{N_{\rm s}} { (\frac{\mu_j^{\rm s}-\mu_{j}^{\rm star}}{\delta \mu_{j}^{\rm s} })^2} + \sum_{j=1}^{N_{\rm s}} { (\frac{\sigma_j^{\rm s}-\sigma_{j}^{\rm star}}{\delta \sigma_{j}^{\rm s} })^2} ,
\end{equation}
where $N_{\rm s}$ is the number of bins in the integrated-light stellar kinematic data,  and $(\mu_{j}^{\rm star}, \sigma_{j}^{\rm star})$ are the model predicted mean velocity and velocity dispersion for the stellar tracer -- which we also used for the PNe.

The minimum of $\chi^2_{\rm star}$ and the maximum of the log likelihood for the discrete stars can be determined at the same time by maximising the combined likelihood defined as 
\begin{equation}
\label{eqn:Ltot}
\mathcal{L}_{\rm tot} = \mathcal{L} - \frac{1}{2} \alpha_{\rm s} \chi^2_{\rm star}
\\
= \mathcal{L}_{\rm GC} +  \mathcal{L}_{\rm PN} - \frac{1}{2} \alpha_s \chi^2_{\rm star},
\end{equation}
where $\alpha_{\rm s}$ is a weight parameter which will be adapted manually to balance the relative influence of the integrated stellar kinematic data and the GC and PN discrete data (see Section~\ref{SS:modelparam}). 

\subsection{Model parameters and optimization}
\label{SS:modelparam}
Using the prescriptions laid out in Section~\ref{SS:potential}, there are three free parameters in the gravitational potential:
\begin{itemize}
\item[(1)] $\Upsilon_V$, the $V$-band stellar mass-to-light ratio;
\item[(2)] $\rho_s$, DM scale density;
\item[(3)]  $d_s \equiv \log (\rho_s^2 r_s^3)$ is a proxy for the scale radius $r_s$, reducing the strong degeneracy between $\rho_s$ and $r_s$. 
\end{itemize}

We have two GC populations, red and blue, with different chemical, spatial and dynamical distributions.  The projected number density of each population is fixed, as shown in Fig.~\ref{fig:sdsb}, leaving two chemical and two dynamical free parameters for each population, and thus 
there are four free parameters for the red population:
\begin{itemize}
\item[(4)]  $Z_0^\mathrm{red}$, mean of the Gaussian colour distribution;
\item[(5)] $\sigma_Z^\mathrm{red}$, dispersion of the Gaussian colour distribution;
\item[(6)] $\lambda^\mathrm{red}$, $\equiv - \ln \left( 1 - \beta^\mathrm{red}_z \right)$, a symmetric recasting of the constant velocity anisotropy in the meridional plane $\beta^\mathrm{red}_z$;
\item[(7)] $\kappa^\mathrm{red}$, rotation parameter.
\end{itemize}

There are correspondingly four free parameters for the blue population:
\begin{itemize}
\item[(8)]  $Z_0^\mathrm{blue}$;
\item[(9)] $\sigma_Z^\mathrm{blue}$;
\item[(10)] $\lambda^\mathrm{blue}$; 
\item[(11)] $\kappa^\mathrm{blue} $.
\end{itemize}

The PNe and stars are treated together as stellar tracers with chemical properties decoupled from the GCs. The projected stellar density profile is set by the surface brightness profile in Fig.~\ref{fig:sdsb}, and there are two free dynamical parameters left to be defined:
\begin{itemize}
\item[(12)] $\lambda^\mathrm{star}$, $\equiv - \ln \left( 1 - \beta^\mathrm{star}_z \right)$, the velocity anisotropy parameter of the stars/PNe;
\item[(13)] $\kappa^\mathrm{star} $, the rotation parameter of the stars/PNe.
\end{itemize}

In order to understand the ability of our model to distinguish between different DM haloes, we run two sets of models: one set using cored DM haloes and one set using cusped DM haloes. For each model, we have 13 free parameters as described above.  

\begin{figure*}
\centering\includegraphics[width=\hsize]{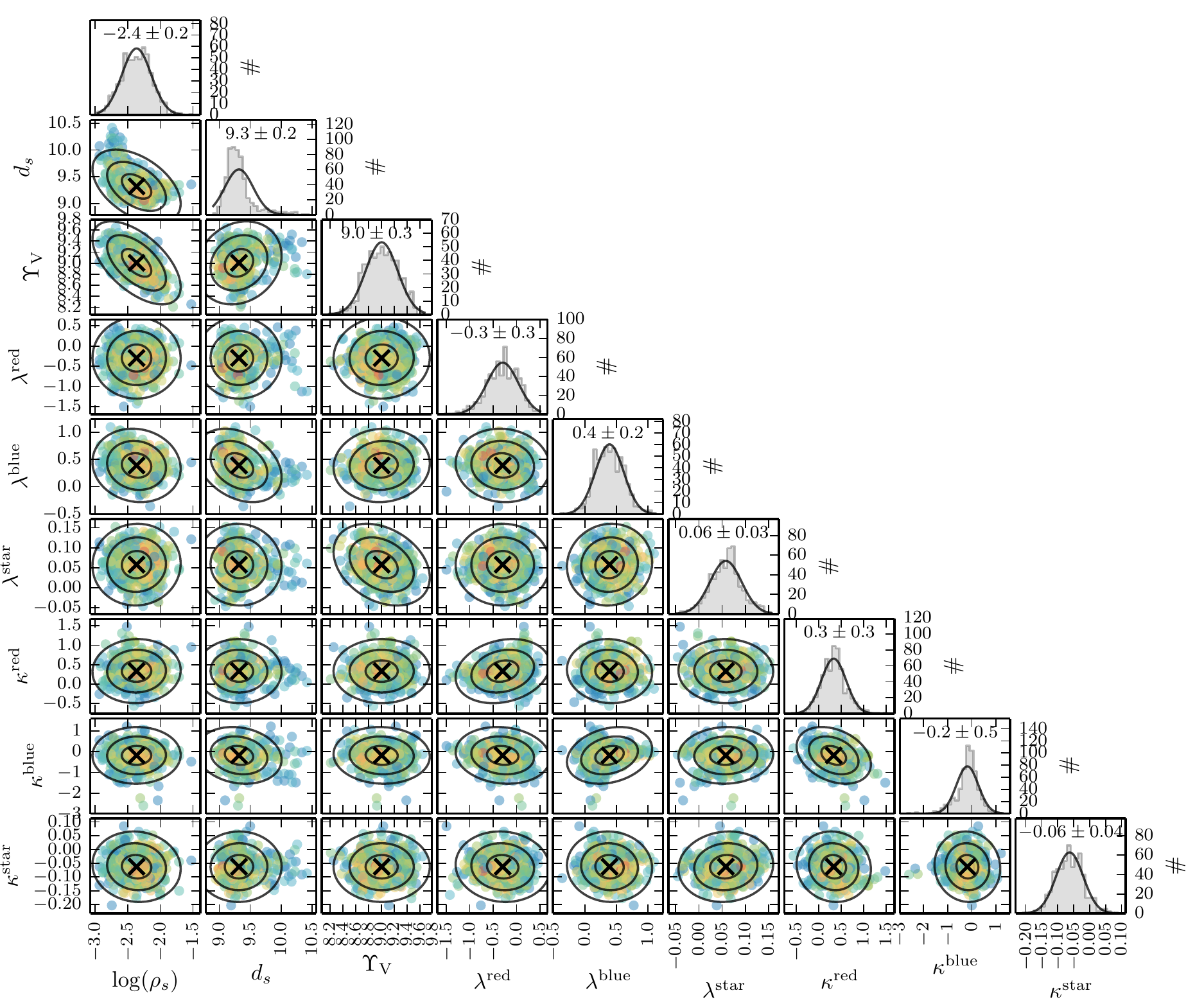}
\caption{MCMC post-burn distributions for the cored DM halo ($\eta = 2, \gamma =0$) model. The scatter plots show the projected 2D distributions, with the points coloured by their likelihoods from blue (low) to red (high). The ellipses represent the $1\sigma$, $2\sigma$ and $3\sigma$ regions of the projected convariance matrix. The histograms show the projected 1D distributions, with curves representing the $1\sigma$ projected covariance matrix.  Note that we only show 9 out of 13 parameters here: from left to right, they are the logarithm of the DM scale density $\log \rho_s$; $d_s = \rho_s^2 r_s^3$ where $r_s$ is the DM scale radius; the stellar mass-to-light ratio $\Upsilon_V$; the velocity anisotropy parameters for the red GCs $\lambda^\mathrm{red}$, the blue GCs $\lambda^\mathrm{blue}$ and the stellar tracers $\lambda^\mathrm{star}$;  the rotation parameters for the red GCs $\kappa^\mathrm{red}$, the blue GCs $\kappa^\mathrm{blue}$ and the stellar tracers $\kappa^\mathrm{star}$. }
\label{fig:mcmc_postburn}
\end{figure*}

The \textsc{emcee} package \citep{Foreman-Mackey2013}, a Python implementation of the an affine-invariant Markov Chain Monte Carlo (MCMC) ensemble sampler, is used to run the models. For each set of models, 200 walkers with 300 steps are employed, where walkers are the members of the ensemble. The walkers are similar to separate Metropolis--Hasting chains but the proposal distribution for a given walker depends on the positions of all the other walkers in the ensemble. We burn in the chain at the time step 250, only the last 50 steps are used for the post-burn distributions as shown in Fig.~\ref{fig:mcmc_postburn}.

Parameter $\alpha_{\rm s}$ in equation~(\ref{eqn:Ltot}) is used to numerically balance the equation and set the relative influence of the discrete data extending to large scales and the integrated stellar kinematic data at small scales. We manually vary $\alpha_{\rm s}$ and find that, at small values, the integrated-light stellar kinematic data provides a weak constraint on the model, with large variations in $\Upsilon_V$, and hence in the DM mass.  When $\alpha_{\rm s}$ is large, the integrated stellar kinematic data dominate the evolution of the log likelihood, and cause a poor fit to the discrete data.  For our NGC 5846 model, the SAURON IFU data have about $1000$ good quality data points and so we find that a small value,  $\alpha_{\rm s} = 0.1$, leads to good fits to both the integrated stellar kinematic data and the discrete PNe velocities.  


\section{Model fits}
\label{S:modelfits}

The best-fitting parameters obtained by the MCMC process for the models with cored and cusped DM potentials are presented in Table~\ref{tab:para} (with $d_s$ converted to the scale radius $r_s$, and $\lambda^\mathrm{red}$, $\lambda^\mathrm{blue}$ and $\lambda^\mathrm{star}$ converted to the velocity anisotropy parameters $\beta_z^\mathrm{red}$, $\beta_z^\mathrm{blue}$ and $\beta_z^\mathrm{star}$ for convenience). 

Fig.~\ref{fig:mcmc_postburn} shows the resulting distributions for 9 of the 13 free parameters for the models with a cored DM halo; the 4 parameters related to the colour distribution of the GCs are omitted as they are not directly related to the dynamics. The scatter plots show the projected 2D distributions, with the points coloured by their likelihoods from blue (low) to red (high). The ellipses represent the $1\sigma$, $2\sigma$ and $3\sigma$ regions of the projected covariance matrix. The histograms show the projected 1D distributions, with bell-like curves representing the $1\sigma$ projected covariance matrix.   
In most cases there are no degeneracies between the parameters, except for some covariance between
the stellar mass-to-light ratio $\Upsilon_V$, stellar velocity anisotropy $\lambda^{\mathrm{star}}$, and DM scale density $\rho_s$ (representating classic mass--anisotropy and mass decomposition
degeneracies). Unlike the spherical Jeans model, the mass--anisotropy degeneracy is not strong in our model.
The MCMC process works similarly for the cusped model and the corresponding parameters show similar degeneracies. 

We calculate the virial mass $M_{200}$ and the concentration $c$ of the DM halo. The virial mass $M_{200}$ is defined as the enclosed mass within the virial radius $r_v$, the virial radius $r_v$ taken as $r_{200}$, where the average density inside $r_{200}$ is 200 times the critical density ($\rho_{\mathrm{crit}} = 1.37 \times 10^{-7} \,M_{\odot}\,\mathrm{pc}^{-3}$). The concentration $c$ is defined as the ratio between the DM virial radius $r_v$ and the DM scale radius $r_s$.  
$M_{200}$ and $c$ are shown in Table~\ref{tab:para} as well as the scale density $\rho_s$ and the scale radius $r_s$. 
The cusped DM halo has a remarkably low concentration and large virial mass, which is not expected from cosmological simulations
(and discussed in Section~\ref{S:mass}).

We run a constrained cusped (CC) model with $\rho_s > 1\times10^{-3} M_{\odot}$~pc$^{-3}$ (this roughly corresponds to a standard NFW halo as a prior) to see statistically how strongly we can constrain the concentration of the halo.
The best-fitting parameters of this constrained cusped model are also listed in Table~\ref{tab:para}. 

The inclination angle is fixed at $90^\circ$. We have also investigated the effect of leaving the inclination angle as a free parameter. We find that it is poorly constrained in the range from $50^\circ$ to $90^\circ$. The rotation parameter of the red GCs does increase slightly, but otherwise a free inclination angle does not have a significant effect on our results. 

\begin{table*}
\caption{The best-fit model parameters obtained by the MCMC process with the cored, cusped and the constrained cusped (CC; with $\rho_s > 1\times 10^{-3}\, M_{\sun} \, \mathrm{pc}^{-3}$ as a prior) DM haloes. The parameters are presented in two rows for each model. The first row from left to right: DM scale density $\rho_s \, [10^{-3}M_{\odot}\mathrm{pc}^{-3}] $, the DM concentration $c$, DM scale radius $r_s$~[kpc], the logarithm of the virial mass $\log {M_{\rm vir}\, [M_{\odot}]}$, the $V$-band stellar mass-to-light ratio $\Upsilon_V$, the means of the Gaussian $g-i$ colour distributions of the red GCs $Z_0^\mathrm{red}$ and the blue GCs $Z_0^\mathrm{blue}$, the dispersion of the Gaussians of the red GCs $\sigma_Z^\mathrm{red}$ and the blue GCs $\sigma_Z^\mathrm{blue}$.
The second row from left to right:  velocity anisotropy in the meridional plane for the stellar tracers $\beta^\mathrm{star}_z$ , the red GCs $\beta^\mathrm{red}_z$  and the blue GCs $\beta^\mathrm{blue}_z$,  the rotation parameters $\kappa$ of the stellar tracers, the red GCs and the blue GCs, $\mathcal{L}_{\rm max}$ which is the maximum likelihood achieved for that set of models and $\delta \mathcal{L}$ for the $1\sigma$ confidence level. 
Note that the directly fitted parameter $d_s$ has been converted to $r_s$, and $\lambda^\mathrm{star}$, $\lambda^\mathrm{red}$ and $\lambda^\mathrm{blue}$ have been converted to $\beta_z^\mathrm{star}$, $\beta_z^\mathrm{red}$ and $\beta_z^\mathrm{blue}$.}

\label{tab:para}
\small
\scriptsize
\footnotesize
\begin{tabular}{llllllllll}
\hline
\hline
 DM & $\rho_s $ & $c$ & $r_s $ & $\log{M_{\rm vir}}$ &  $\Upsilon_V$ &  $Z_0^\mathrm{red}$   & $Z_0^\mathrm{blue}$ & $\sigma_Z^\mathrm{red}$ & $\sigma_Z^\mathrm{blue}$\\
\\
  &  $\beta_z^\mathrm{star}$  & $\beta_z^\mathrm{red}$ &  $\beta_z^\mathrm{blue}$  &  $\kappa ^{\rm star}$&   $\kappa ^{\rm red}$   & $\kappa ^{\rm blue}$ & & $\mathcal{L}_{\rm max}$  & $\delta L (1\sigma)$ \\
 \hline
  \hline
 
Cored &  $4.0^{+2.3}_{-1.5}$      &  -           &  $50^{+30}_{-20} $         & $13.1^{+0.7}_{-0.9}$   & $9.0\pm0.3$                &   $1.09\pm0.02$   & $0.87\pm0.02$    & $0.15\pm0.02$  & $0.12\pm0.02$  \\
           & $ 0.05^{+0.05}_{-0.05}$  &  $-0.34^{+0.32}_{-0.50}$   & $0.32^{+0.13}_{-0.16}$  &   $-0.07\pm0.04$  & $0.3\pm0.2$       & $-0.1\pm0.4$    & & $-7415$ & 6 \\
 \hline
Cusped &  $0.025^{+0.17}_{-0.022}$  &  $1.2^{+1.2}_{-0.6}$ & $1149^{+6000}_{-600} $  & $15.4^{+2.4}_{-2.6}$ &  $8.7\pm0.3$                &   $1.10\pm0.02$   & $0.86\pm0.02$    & $0.15\pm0.02$  & $0.12\pm0.02$  \\
           & $ 0.05^{+0.04}_{-0.05}$  &  $-0.34^{+0.36}_{-0.50}$   & $0.35^{+0.13}_{-0.17}$  &   $0.08\pm0.01$  & $0.3\pm0.2$       & $-0.1\pm0.3$  & &  $-7415$ & 7 \\
\hline

CC &  $1.6^{+0.9}_{-0.6}$  & $5.7_{-0.7}^{+1.2}$  & $74^{+37}_{-23} $ & $12.9^{+0.8}_{-0.4}$& $8.3\pm0.3$                &   $1.10\pm0.02$   & $0.86\pm0.02$    & $0.15\pm0.02$  & $0.12\pm0.02$  \\
           & $ 0.08^{+0.01}_{-0.01}$  &  $-0.49^{+0.36}_{-0.50}$   & $0.33^{+0.13}_{-0.17}$  &   $0.08\pm0.01$  & $0.3\pm0.2$       & $-0.1\pm0.3$   & & $-7417$ & 6 \\

  \hline
  \hline
 \end{tabular}
\end{table*}

\subsection{The chemo-dynamical GCs}
\label{SS:chemo}
\begin{figure}
\centering\includegraphics[width=\hsize]{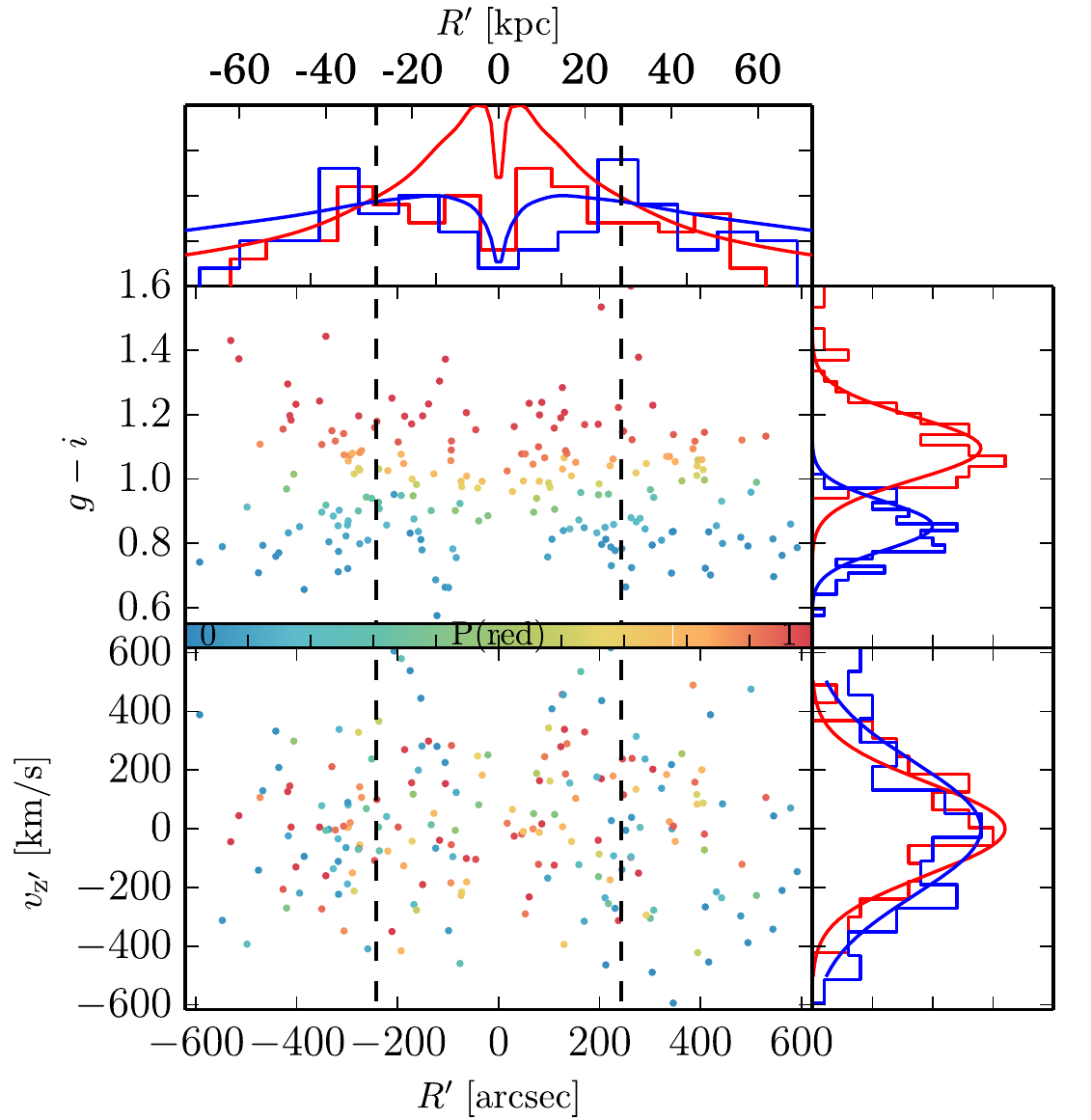}
\caption{Chemo-dynamical modelling results for the spatial, colour and kinematic distributions of GCs in NGC 5846. {\bf Top scatter panel}: The projected semi-major elliptical shell radius $R'$ vs.\ the colour $g-i$.  {\bf Bottom scatter panel}: $R'$ vs. 
line-of-sight velocity $v_{z'}$. GCs are plotted with points coloured by $P_i^\mathrm{'red}$-- the probability of belonging to the red population--from blue (low) to red (high).  
The red and blue histograms show the distributions in radius, color and velocity for the GCs classified as red and blue. The solid curves over-plotted are the model distributions for each population. 
The radial distributions, inferred from the surface number density profiles, do not match the histograms in the inner regions owing to spectroscopic incompleteness.
Note also that the velocity dispersions actually vary with radius, but here we plot the overall velocity distribution for each population. The two vertical dashed lines indicate the positions of $-3\,R_{\rm e}$ and $3\,R_{\rm e}$. }
\label{fig:chem-dyn}
\end{figure}

The key approach in our chemo-dynamical modelling is that the 213 GCs with velocity measurements are not separated into red and blue groups before performing the dynamical modelling. 
Instead, from the best model obtained with the MCMC process, we calculate the probabilities of an individual GC being red ($P_i^\mathrm{'red}$) or blue ($P_i^\mathrm{'blue}$ ) following equation~(\ref{eqn:Pik}). 
We show the chemo-dynamical separation of the GCs in our best-fitting cored model in Fig.~\ref{fig:chem-dyn}. The results for the cusped DM halo and the constrained DM halo models are very similar, so we do not include them here. 
The upper and lower scatter plots show the distribution of GCs in colour $g-i$, and the line-of-sight velocity $v_{z'}$ versus the projected semi-major elliptical shell radius $R'$ defined in equation~(\ref{eqn:RR}). 
The GCs are plotted with points coloured by $P_i^\mathrm{'red}$:
 the red points representing GCs with high probability of being in the red population, and the blue points representing GCs with high probability of being in the blue population. 
The red and blue histograms are constructed directly from the GCs identified by $P_i^\mathrm{'red} > 0.5$ for red and $P_i^\mathrm{'blue} > 0.5$ for blue. This criterion for separation is used in all the following sections of the paper.
The solid curves in the histograms are the best-fit model-predicted distributions for each population. 
The model-predicted radial distributions $f(R')$ are inferred from the photometric surface number density profiles $\Sigma(R)$, with $f(R') = f(R)/2 = \pi qR\Sigma(R)dR$, where $R'$ is the semi-major elliptical annular radius defined in Equation~\ref{eqn:RR}.

The model input surface number density profiles of the red and blue GCs were determined by photometrically-detected GCs with separation at $g-i < 0.90$ for blue and $g-i> 0.95$ for red (Fig.~\ref{fig:sdsb}).
In the final colour distribution obtained by the chemo-dynamical model in Fig.~\ref{fig:chem-dyn},  the colour distributions of the two GC populations with velocity 
measurements are clearly separated with a small overlap around $g-i \sim 0.9-1.0$.
These distributions are reasonably consistent with the results from a standard, non-dynamical bimodal colour modelling approach \citep{Napolitano2014}.
The surface number density profiles used for the modelling were created in a way that turns out to be nicely consistent with the separation inferred from the dynamical tracers. 

\subsection{Fit to the kinematic data}
\label{SS:fit}
\begin{figure*}
\centering\includegraphics[width=\hsize]{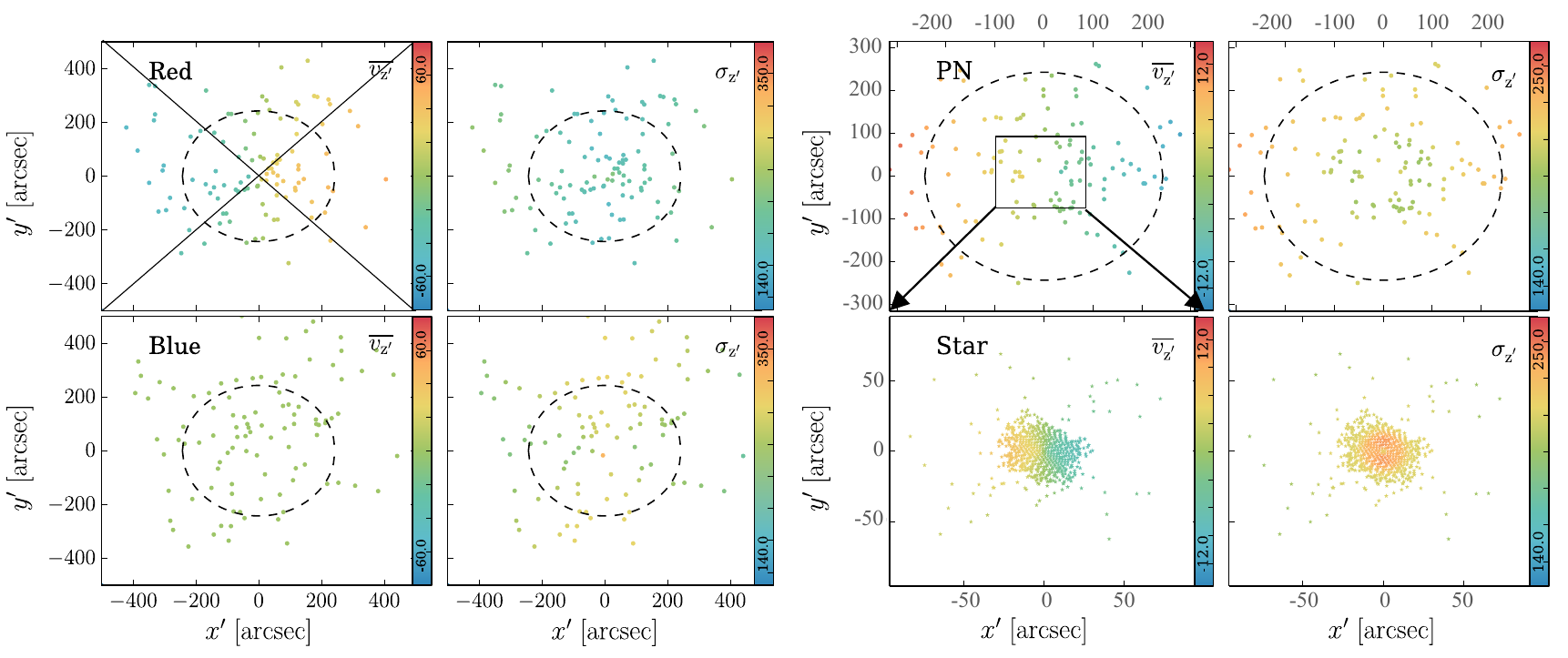}
\caption{The model predicted kinematics of the multiple tracer populations in NGC 5846 with the best-fitting cored dark matter model. The left panels are for globular clusters (GCs) and the right panels are for planetary nebulae (PNe) and stars. 
{\bf Left:} The best-fitting model predicted mean velocity and velocity dispersion maps for the red GCs (top) and blue GCs (bottom), where each point represents a GC position coloured with the predicted value, which scales as indicated by the corresponding colour bar.
{\bf Right:} The best-fit model predicted mean velocity and velocity dispersion maps for the stellar tracers in the regions covered by
the kinematic data from the PNe and from the integrated stellar light (SAURON and SLUGGS), where the points are coloured on the same scale as the GCs. 
Note that the PN data extend to $\sim 300''$ and the integrated stellar data to $\sim 100''$.  
The dashed circles mark the $3\,R_{\rm e}$ radius.
 }
\label{fig:kin_map}
\end{figure*}

\begin{figure*}
\centering\includegraphics[width=18cm, height = 10cm]{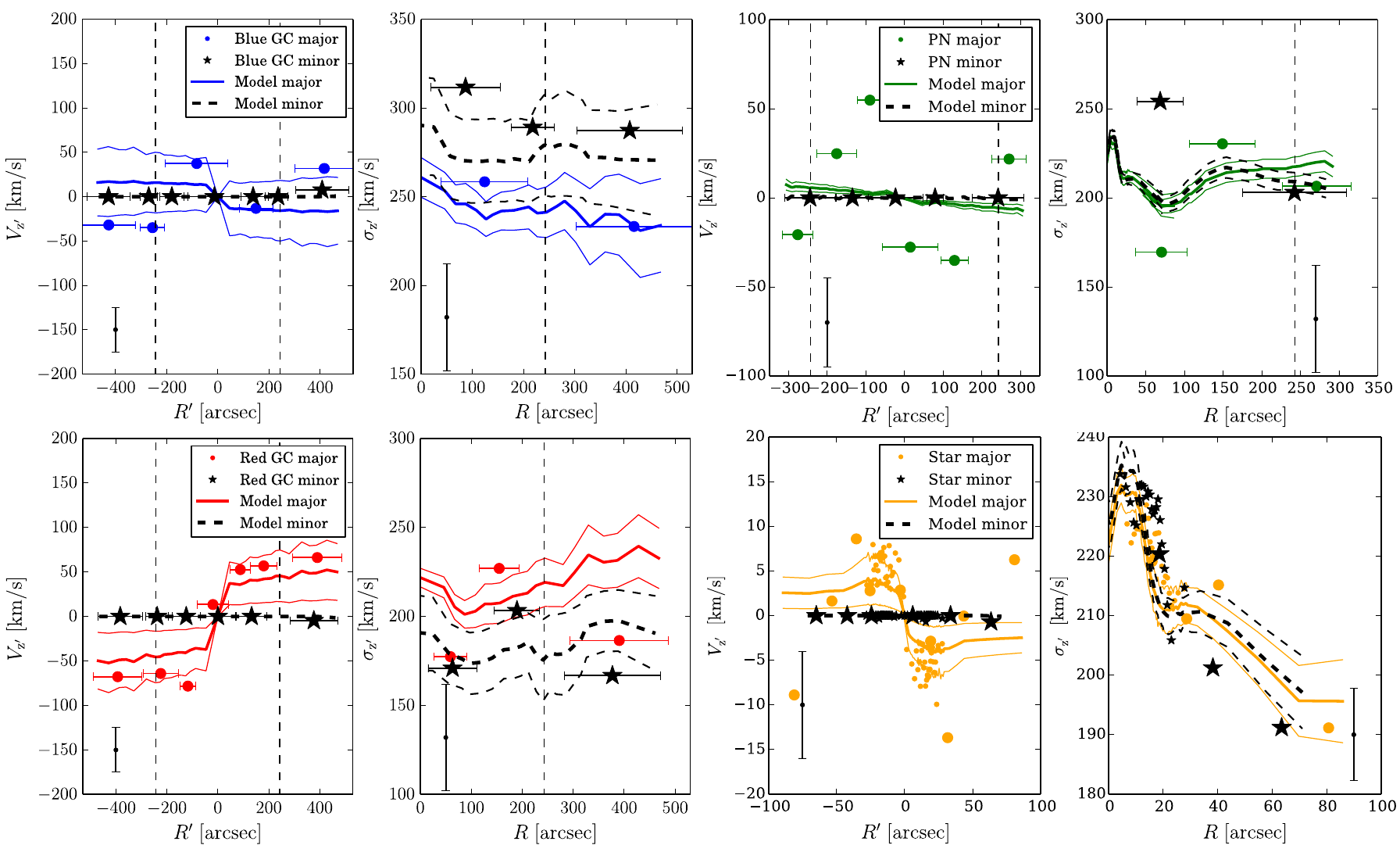}
\caption{Comparison between the data and the model predicted kinematics from the best-fitting cored dark matter model. The left panels are for globular clusters (GCs) and the right panels are for planetary nebulae (PNe) and stars. 
{\bf Left:} The mean velocity and velocity dispersion profiles for the red (top) and the blue (bottom) GCs. 
The data and model predictions were binned along the major and minor axes using the cone-like divisions as shown in Fig~\ref{fig:kin_map}. 
The thick and thin curves represent the mean and $1\sigma$ uncertainties of the model inferences for the GCs; the solid and dashed lines represent those along the minor and minor axis, respectively.
The points with error bars are the actual observational data, in radial bins, with red and blue GCs classified as such by the best-fit likelihood model.
The points and star-symbols are for the major and minor axes, respectively.
{\bf Right:} The binned mean velocity and velocity dispersion profiles, where the solid and dashed curves represent those of the model inferences for the field stars about major and minor axes. 
The points and star-symbols are the data binned along the major and minor axes, respectively. In the right bottom panel, the large symbols are binned from the SLUGGS data and the small symbols represent the SAURON data. 
The typical uncertainty for the binned data points is indicated on the bottom left of each panel.  
The vertical lines mark the $3\,R_{\rm e}$ radius.
 }
\label{fig:kin}
\end{figure*}

The kinematic maps from the best-fitting cored model are shown in Fig.~\ref{fig:kin_map}, the left panels are the kinematic maps for the red GCs and the blue GCs; while the right panels are for the stellar tracers corresponding to the PNe and integrated-light stellar kinematic measurements.
The kinematic maps are the predicted mean velocity (left columns) and velocity dispersion (right columns) in the projected plane, with each point representing a measurement position colour-coded with the predicted value (as indicated by the corresponding colour bar). The velocity dispersion anisotropy is encoded in the opening angle of the kinematic maps (\citealt{Li2016}, \citealt{Cappellari2008}). Red GCs have higher dispersions along the major axis while the blue GCs have higher dispersions along the minor axis, indicating their different velocity anisotropies. We will discuss this further in Section~\ref{SS:anisotropy}.

We bin the data and the model predictions to show a direct comparison in Fig.~\ref{fig:kin}. 
We divide the projected plane into two cones, one spanning a $\pm45$ deg from the major axis, and the other a $\pm45$ deg from the minor axis as indicated on the top-left panel. The data and model predictions are binned along the major and minor axes using points in the corresponding cones. Both the data and model are axis-symmetrized and point-symmetrized to increase the number of points by a factor of 4 before binning. 

Fig~\ref{fig:kin} shows the binned mean velocity and velocity dispersion profiles.
In each panel, the coloured points are constructed from the observational data along the major axis, while the black stars are along the minor axis. The coloured solid curves are the mean and $1\sigma$ uncertainties of the model predictions along the major axis, and the black dashed curves are along the minor axis. 
 
Equally-populated radial bins are used for the GC and PN discrete data with 30 (40 for PN) points per bin counting from left to right. Adjacent bins are independent.
The mid-way position from the minimum to maximum $R'$ of the 30 points is taken as the $R'$ value of a bin. The horizontal error bar covers the $R'$ region that the 30 points span. For the stellar kinematic data, we omit the horizontal error bars, while the SAURON data are binned 40 points per bin and the SLUGGS data are binned with 20 points per bin, with adjacent bins independent from each other. 

The data points fluctuate substantially, with the typical uncertainty shown in each panel. 
The observed rotation in the red GCs and in the integrated starlight in the inner regions is significant and matched well by the model prediction. 
The stellar kinematics exhibit a weak kinematically decoupled core (KDC) inside $\sim$~30~arcsec 
which is not matched well by our model.
The observed PNe mean velocity profile has large fluctuations and shows no obvious rotation, which is consistent with our model predicts, agrees with the result from \citet{Coccato2009}. The rotation of these different tracers will be discussed further in Section~\ref{SS:rot}.

The model-predicted velocity dispersion profiles for all the tracers
match the observational data reasonably well. The dispersion profile is matched radially, and is the velocity dispersion difference between the major and minor axes. 

The cusped model fits the data as well as the cored model. The constrained cusped model provides a poorer fit to the stellar kinematics, with a too large velocity dispersion around $20-100$ arcsec predicted by the model. The kinematics of the cusped model and the constrained cusped model are shown in Appendix~\ref{S:kin_c5}. 
\section{Kinematic properties of the three tracers}
\label{S:kin}
\subsection{Internal rotation}
\label{SS:rot}
\begin{figure*}
\centering\includegraphics[width=\hsize]{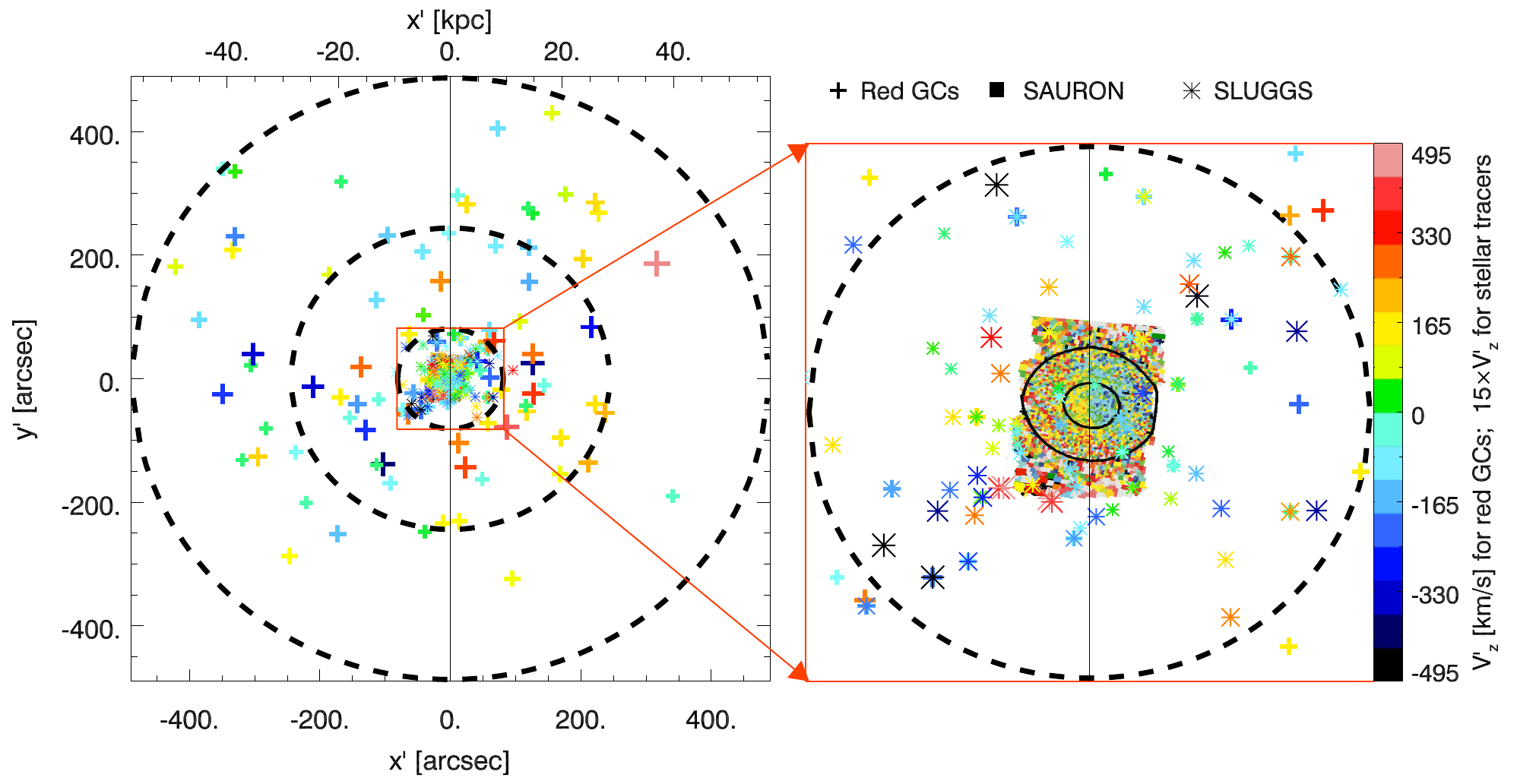}
\caption{The positions of the tracers coloured by their measured line-of-sight (LOS) velocity. The plus-symbols represent the red GCs (as identified by the model), the asterisks are the SLUGGS stellar kinematics data, and the central velocity map showing Voronoi bins displays the SAURON data. Note that the velocity scale for the red GCs is 15 times that of the field stars. The sizes of the symbols also scales with the absolute value of the LOS velocity. The three dashed circles indicate $1\,R_{\rm e}$, $3\,R_{\rm e}$ and $6\,R_{\rm e}$ of the 
galaxy light, we zoom in to the inner $1\,R_{\rm e}$ region in the right panel, the two solid ellipses are contours of surface brightness of the galaxy in the inner region. The vertical line indicates the photometric minor axis of the galaxy.}
\label{fig:xy_rot}
\end{figure*}

\begin{figure}
\centering\includegraphics[width=\hsize]{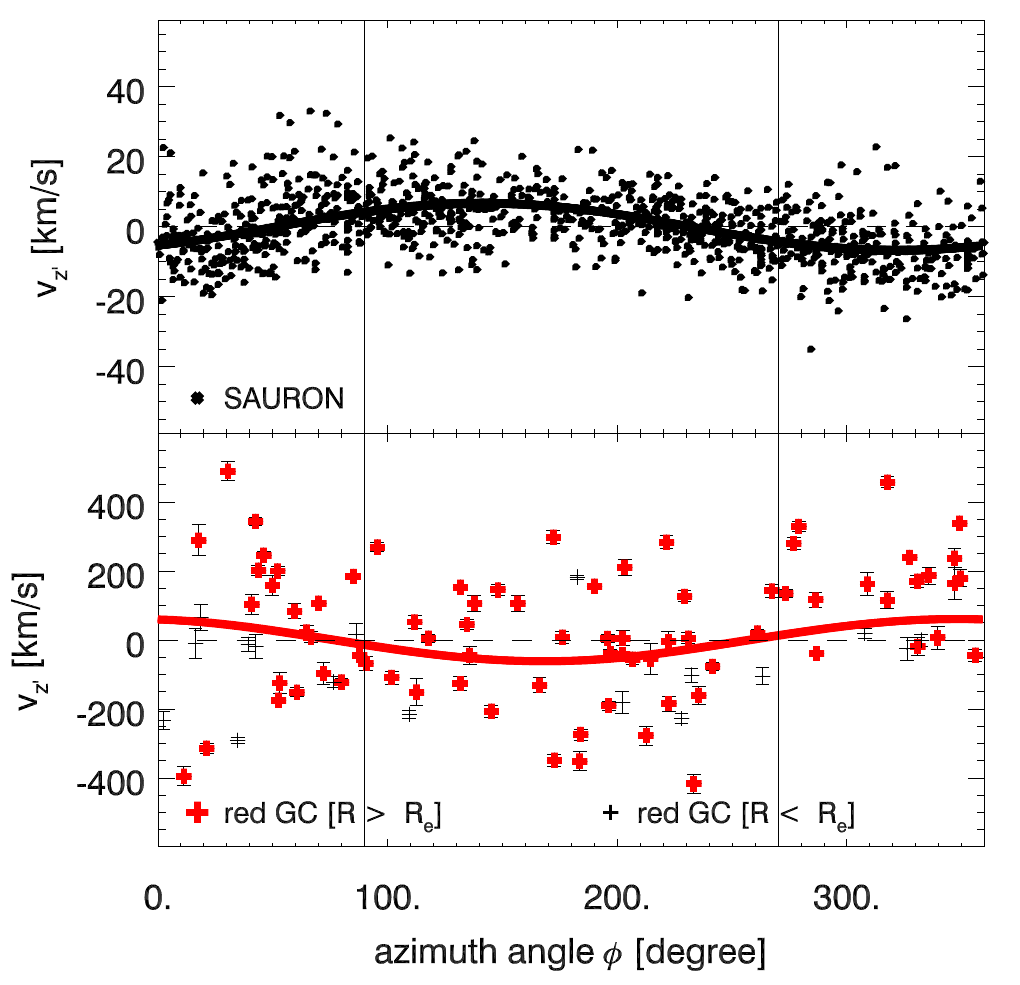}
\caption{ The variation of line-of-sight velocity $v_{z'}$ along the azimuthal angle ($\phi$) measured from the major axis of the galaxy. 
{\bf Top panel:} The points are the SAURON data, and the black curve is the sinusoidal fit to the data. 
{\bf Bottom panel:} The thin black pluses represent the red GCs at $R< R_{\rm e}$, and the red thick pluses are the red GCs at $R>R_{\rm e}$, with the red curve as the corresponding sinusoidal fit. }
\label{fig:rot_fit}
\end{figure}

We identify the red GCs based on their likelihoods in the best-fitting model.
As shown in Figure~\ref{fig:kin}, the field stars and red GCs show significant rotation. In our axisymmetric model, only the rotation about the minor axis is fitted by our modelling: see the complicated rotation pattern shown in Figure~\ref{fig:kin}.  
As we discuss below, the PNe and the blue GCs are consistent with no rotation, so we only show the red GCs and the field stars in the following two figures. 

Figure~\ref{fig:xy_rot} shows the positions of the tracers coloured by their LOS velocities, illustrating the KDC in the inner region. The velocity gradient of the field stars traced by the SLUGGS data (asterisks) at $\sim 0.5$--1.0\, $R_{\rm e}$ starts differ from that in the inner $0.5\,R_{\rm e}$ as traced by the SAURON data (the velocity map showing the Voronoi binning). Further out, the red GCs (plus symbols) show rotation in the opposite direction from the SAURON data,
and thus the region traced by the SLUGGS stellar kinematics is likely to be at the transition of these counter-rotating inner and outer regions. 

The rotation directions of the SAURON data and the red GCs are roughly around the photometric minor axis of the galaxy. The variation of LOS velocities along the azimuthal angle $\phi$ measured from the major axis of the galaxy is shown in Fig.~\ref{fig:rot_fit}. 
A simple sinusoidal fit to the SAURON data,
$v = v_{\rm max} \sin(\phi + \phi_0)$,
 yields $\phi_0 = -60^\circ \pm 20^\circ$ and $v_{\rm max} = 7.0\pm 2.2$ \kms , as indicated by the solid black curve in the top panel.
The sinusoidal fit to the red GCs at $R > \,R_{\rm e}$ yields the red solid curves with $\phi_0 = 93^\circ \pm 6^\circ$ and $v_{\rm max} = 61 \pm 21$ \kms,
with hints of the rotation increasing with radius.
For the red GCs with $\sigma^\mathrm{red} \approx 200$ \kms , we get $v_{\rm max}/\sigma \sim 0.3$ at $R>R_{\rm e}$. 
At $R < R_{\rm e}$, there are only a few red GCs, and no obvious rotation pattern.
These fits confirm that the central galaxy light and the red GCs both have rotation around the minor axis (along the major axis) but in opposite directions.

We also tried the sinusoidal fit to the PNe and blue GCs, both of which are consistent with zero rotation, as well as that suggested by the dynamical model fitting as shown in the last section.
The PNe are, if anything, counter-rotating to the red GCs at the same radii. 
This raises the possibility that the red GCs do not surprisingly trace the field stars (PNe) spatial distribution - as noted earlier. 
The blue GCs are also consistent with no rotation at all radii; they may share the similar rotate properties with the PNe rather than the red GCs.

\subsection{Velocity dispersion anisotropy}
\label{SS:anisotropy}

\begin{figure}
\centering\includegraphics[width=\hsize]{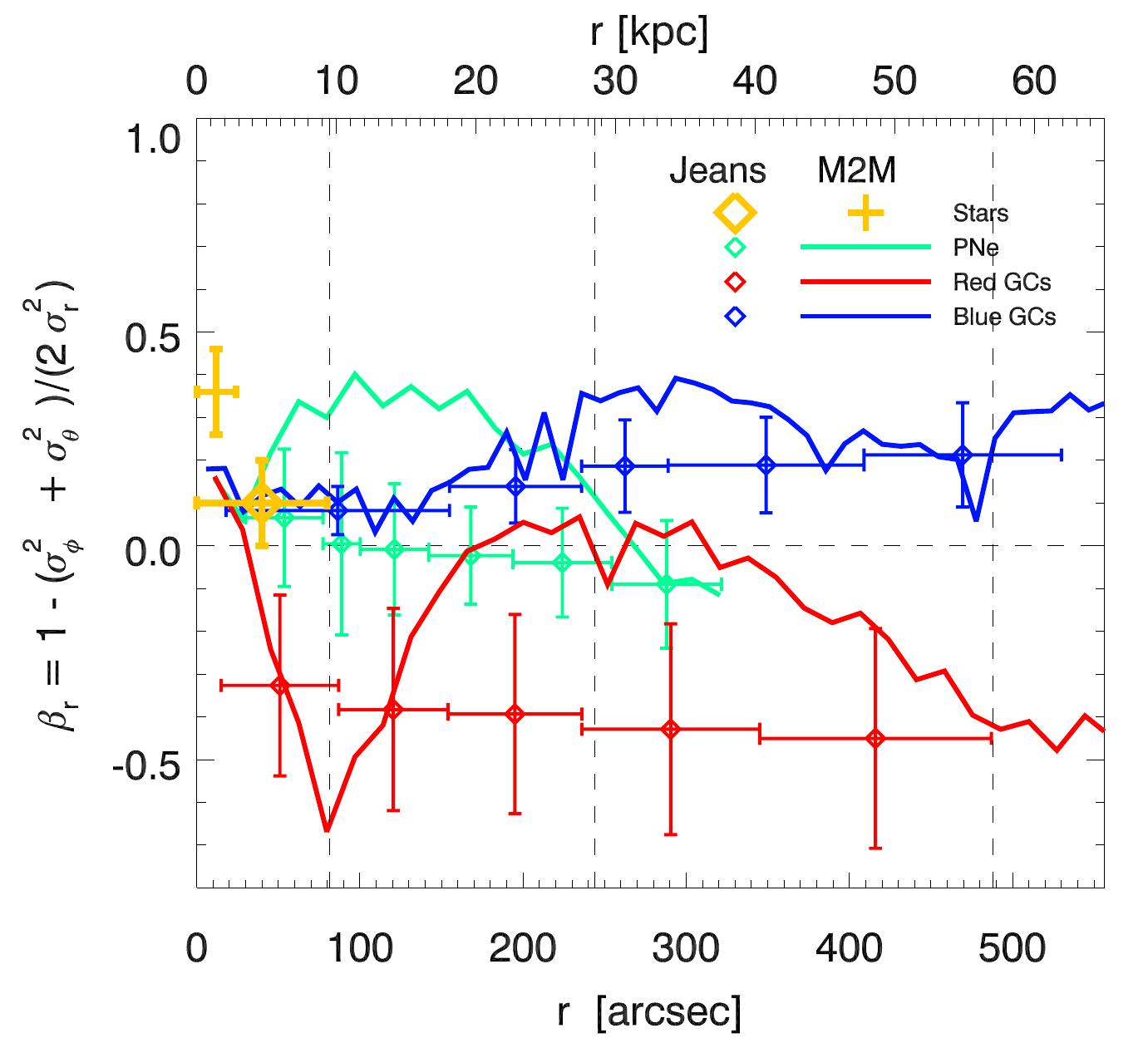}
\caption{The velocity anisotropy profiles in the cored models. The orange, green, red and blue symbols represent field stars, PNe, red GCs and blue GCs.
The diamonds with error bars are the profiles obtained from the discrete chemo-dynamical Jeans model. The solid lines are the results from our M2M models. The large orange plus symbol is the stellar velocity anisotropy obtained by a M2M model from \citet{Long2012}. The three vertical lines represent the positions of $R_{\rm e}$, $3\,R_{\rm e}$ and $6\,R_{\rm e}$. }
\label{fig:beta}
\end{figure}

We calculate the velocity anisotropy profiles from our dynamical models. We use the usual definition for the velocity anisotropy with $\beta_r = 1 - (\sigma_{\phi}^2 + \sigma_{\theta}^2)/2\sigma_{r}^2$ where $\sigma_{\phi}$, $\sigma_{\theta}$ and $\sigma_{r}$ being the velocity dispersion in spherical coordinates, are obtained from the $\sigma_{R}$, $\sigma_{\phi}$ and $\sigma_{z}$ cylindrical values from our axisymmetric models.
The radial profiles for the three populations are shown in Fig~\ref{fig:beta}. 
The profiles are for the models with a cored DM halo, but the results are similar for the cusped halo models. 
The orange, black, red and blue symbols and lines represent the field stars, PNe, red GCs and blue GCs, respectively.

The diamonds with error bars are the velocity anisotropy profiles obtained from the discrete chemo-dynamical Jeans model. 
In the models described in Section~\ref{SS:fit}, the velocity anisotropy of the stellar tracers including PNe is dominated by the stellar kinematic data in the inner region. This is plotted as the orange diamond. The velocity anisotropy in the region covered by the PNe only is plotted with green diamonds.

We find very mild radial velocity anisotropy for the stars in the inner regions.
The PNe match up with the stars, and in the outer parts become mildly tangentially anisotropic.
The blue GCs are slightly radially anisotropic, and the red GCs are tangentially anisotropic.  
Note that these anisotropy profiles are fairly flat by construction, since in our chemo-dynamical model we assumed a constant $\beta_z$ for each population.
   
In an axisymmetric system, radial velocity anisotropy causes the velocity dispersions along the minor axis to be higher, while tangential velocity anisotropy increase the dispersions along the major axis. 
The anisotropy results we obtained for the different populations illustrate this scenario. The blue GCs have higher dispersions along the minor axis (Fig.~\ref{fig:kin})
and they are found to be radially anisotropic. The red GCs at all radii and the outermost PNe have higher dispersions along the major axis and
are they are found to be tangentially anisotropic. 

Having obtained the best-fitting model, the red and blue GCs can be separated via likelihood as described in Section~\ref{SS:chemo} to give three independent discrete tracers: PNe, red GCs and blue GCs. 
To investigate the velocity anisotropies further, we create independent single component particle-based made-to-measure (M2M) models for each population of tracers (\citealt{deLorenzi2007}; \citealt{Long2010}; \citealt{Zhu2014}).  A cored DM halo is adopted with the parameters of the potential fixed by the values obtained by the best-fitting chemo-dynamical Jeans models. The details of the M2M models are described in the Appendix~\ref{S:m2m}.

The M2M models are not used to constrain the potential but to assist with the velocity anisotropy profile analysis, which should provide more robust result by using the information about the opening angle of the kinematic map as well as the whole velocity distribution. 
The solid lines in Fig.~\ref{fig:beta} are the velocity anisotropy profiles obtained from the M2M models.  Again, the red, blue and green lines represent the red GCs, blue GCs and PNe anisotropies. The large orange plus symbol shows the velocity anisotropy of the stars obtained by M2M models using SAURON data alone \citep{Long2012}.

The PNe are found to be radially anisotropic within $3\,R_{\rm e}$, after which the velocity anisotropy starts to decrease, reaching isotropy by the projected radial limit of the data.
The further continuation at larger radii to tangential anisotropy is uncertain.
Red GCs are found to be tangentially anisotropic around $1\, R_{\rm e}$, becoming isotropic around $3\,R_{\rm e}$, after which the velocity anisotropy decreases to become tangentially anisotropic again. 
Blue GCs are found to have radial anisotropy which increases gradually with radius, in agreement with the results from the axisymmetric Jeans model, and in contrast to the profiles for the other tracers.

Overall, the results from Jeans and M2M are reasonably consistent, keeping in mind both the limitations of Jeans models, and 
the possible uncertainties in the M2M models. 
Considering the differences in more detail:
(1) the Jeans models have limited ability to produce radial variations in the velocity anisotropy profiles as seen in the M2M models;
(2) the velocity anisotropy profiles obtained using the Jeans models are, on average, more tangential than those obtained using M2M.
This effect could be caused by deviations of the galaxy's velocity ellipsoid from the zero tilt angles assumed by the axisymmetric Jeans model \citep{Cappellari2008}.

\section{Mass profile constraints}
\label{S:mass}
\begin{figure}
\centering\includegraphics[width=\hsize]{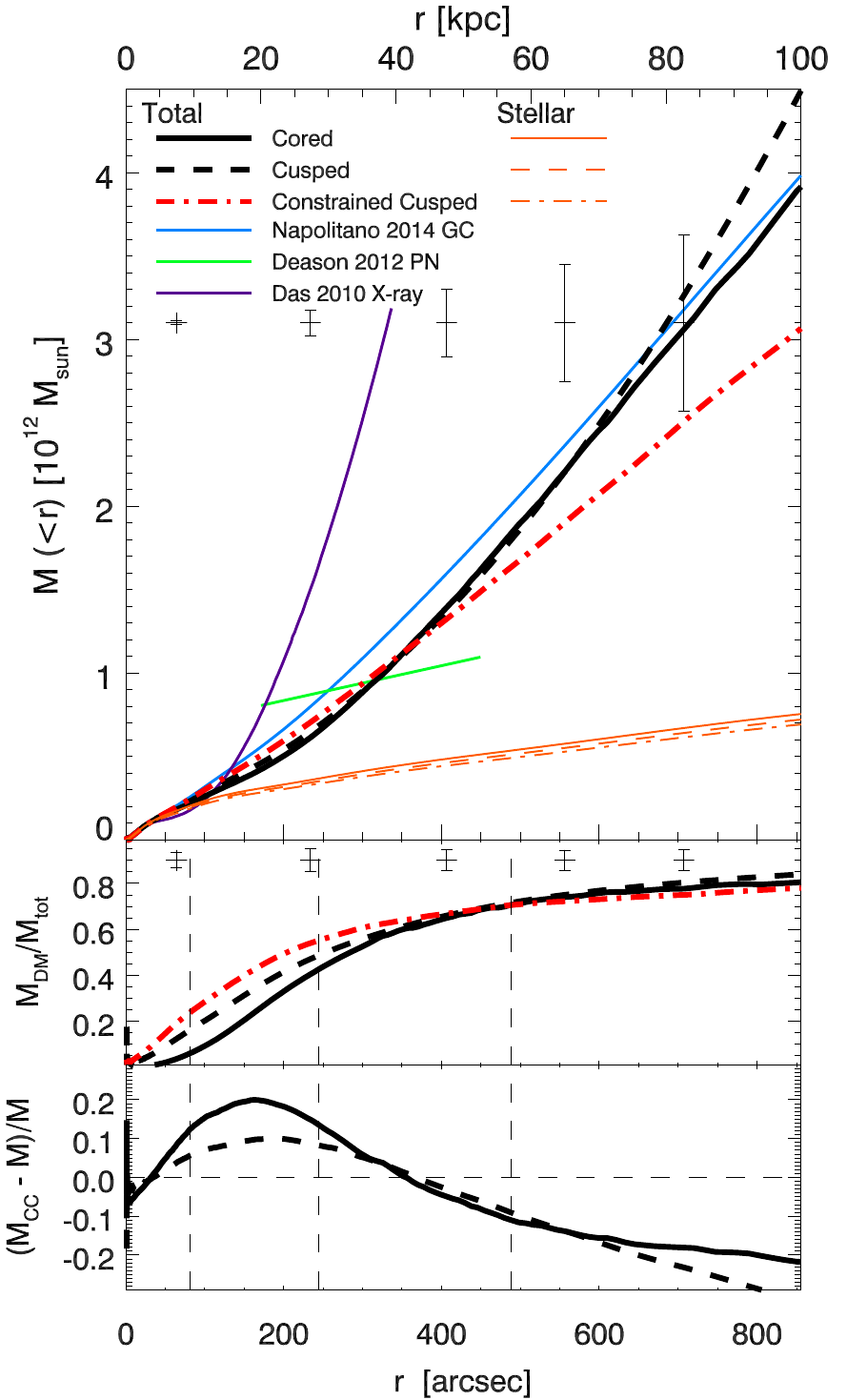}
\caption{{\bf Top panel}: The enclosed mass profiles of NGC 5846.
The black solid, black dashed and red dot-dashed curves represent the total mass profiles of the models with cored, cusped and constrained cusped DM halos, 
while the orange solid, dashed and dot-dashed curves represent the corresponding stellar mass profiles. The pluses with error bars indicate the typical uncertainty in the total mass at those positions.
The blue solid curve represents the standard NFW modelling results from \citet{Napolitano2014}, based on stars and GCs.
The green solid curve is the PN-based mass model from \citet{Deason2012}, and the purple solid curve is the mass profile obtained from modelling the X-ray emission \citep{Das2010}. 
{\bf Middle panel}: The enclosed DM fraction, where the black solid, black dashed and red dot-dashed curves represent the cored DM model, the cusped model and the constrained cusped model. The pluses with error bars indicate the typical uncertainty at those positions.
{\bf Bottom panel}: The relative mass difference of the constrained cusped model compared to the cored model (solid) and the cusped model (dashed). The vertical dashed lines represent the positions of $R_{\rm e}$, $3\,R_{\rm e}$ and $6\,R_{\rm e}$. }
\label{fig:mass}
\end{figure}

The enclosed mass profiles obtained by our best-fitting models are shown in the upper panel of Fig.~\ref{fig:mass}. The middle panel displays the DM fraction profile and the bottom panel, the mass profile difference between the constrained cusped model and the other two models. 

The black solid and dashed curves represent the total mass profile with respectively a cored and a cusped DM halo, the orange solid and dashed curves represent the corresponding stellar mass profiles. 
These two mass profiles are consistent with each other to within $1\sigma$ uncertainty, and they fit all the data with almost equal quality (see Table~\ref{tab:para}).
A stellar mass-to-light ratio of $\Upsilon_V = 9.0 \pm 0.3$ is obtained with the cored DM halo, 
while we estimate $\Upsilon_V = 8.7 \pm 0.1$ with the cusped DM halo (Table~\ref{tab:para}). 
The DM fraction within $1\,R_{\rm e}$ of these two models is only $\sim 10\%$ and $\sim 15\%$ respectively (see lower panel).
By comparison \citealt{Cappellari2013b} found a DM fraction of 14\% through modelling the SAURON data with an axisymmetric Jeans model. 

The cored model yields reasonable values for the DM density around the scale radius and for the virial mass for a giant elliptical galaxy
(Table~\ref{tab:para}; \citealt{Dutton2014}).
In contrast, the cusped halo has a remarkably low density and large scale radius (low concentration and an implausibly large virial mass of $> 10^{15} M_\odot$; see Table~\ref{tab:para}).
This may seem to be evidence against a cusped halo. To investigate this further, we have tried
a constrained cusped halo, using a density prior of $\rho_s > 1\times 10^{-3} \,M_{\odot}\,\mathrm{pc}^{-3}$ which concentrates more DM in the inner regions (red dash-dotted). 
As a consequence of the higher DM concentration, we obtain a lower stellar mass-to-light ratio of $\Upsilon_V = 8.3 \pm 0.3$ (orange dash-dotted lines), and
the DM fraction within $1\,R_{\rm e}$ increases to $\sim 20\%$. As shown in Table~\ref{tab:para}, although this model is not favoured having a smaller maximum likelihood than the previous two models, the statistical significance of the difference is less than $1\,\sigma$. 

The mass profile differences between the constrained cusped model and the other two models are shown in the bottom panel of Fig~\ref{fig:mass}. The solid curve represents the comparison with the cored model and the dashed curve represents the comparison with the cusped model. The differences are most significant in the regions $\sim~2\,R_{\rm e}$ and $\gtrsim 4\,R_{\rm e}$. The velocity dispersion of the PNe and GCs has a typical uncertainty of $\sim 10\%$, so variations in the mass profiles at the $\sim$~10--20\% level are to be expected.
The stellar kinematics around $1\,R_{\rm e}$ ($\sim$~80~arcsec) are most sensitive to this mass profile difference (see Figure~\ref{fig:kin} and Figure~\ref{fig:kin_c5}). More data points around 1--$3\,R_{\rm e}$ or in the outer most regions would improve our ability to distinguish between the cored model and the constrained cusped model. 

As discussed earlier, there is a $\sim$~10~\kms\ offset between the stellar velocity dispersions from SAURON and from SLUGGS.
In our default models, we have shifted the SLUGGS data up to match SAURON, but if we instead move the SAURON dispersions down by 10~\kms, then the 
stellar mass-to-light ratios in the three models decrease by $\Delta \Upsilon_V \sim 0.2$, 
The DM parameters do not change significantly and a cored model is still preferred with similar significance.
If we do not try to match the SLUGGS and SAURON data, then the model fits become slightly worse, as the decreasing stellar velocity dispersions 
from $0-20$~arcsec to $20-100$~arcsec becomes harder to reproduce. This is particularly true for the constrained cusped model which is then excluded with $1\sigma$ significance. 

Mass profile modelling of NGC 5846 has been performed previously 
using stellar kinematics within 
$2\,R_{\rm e}$ (e.g. \citealt{Kronawitter2000,Cappellari2013a}).    
However, tracers extending to larger radii are needed
in order to probe the DM distribution adequately.
\citet{Das2010} derived a mass profile for NGC~5846 using X-ray gas properties from \emph{Chandra} and \emph{XMM-Newton} observations (shown as the purple solid line in Fig.~\ref{fig:mass}). 
This mass is higher than we have found here, probably because of disturbances in the gas that violate the assumption of hydrostatic equilibrium required in the X-ray analysis.
\citet{Deason2012} modelled the PNe dynamics using a power-law distribution function model 
(shown as the green solid line), and obtained results which are consistent with our results over the region covered by most of the PNe. 

\citet{Napolitano2014} utilized a spherical Jeans model of field stars and two GC populations and included velocity distribution kurtosis constraints.
The blue solid curve represents these results obtained with a standard NFW DM model, which is 
in general agreement with our mass profiles. 
The density of their DM halo is higher than our preferred cored model but is and consistent with our constrained cusped model, They found a larger scale radius with the DM mass at all radii larger than our constrained cusped model. 
With a stellar mass-to-light ratio of $\Upsilon_V = 8.2$, they obtained a DM fraction within $1\,R_{\rm e}$ of $\sim 30\%$. 

As we discussed in Section~\ref{SS:fit}, the concentration of the DM halo is sensitive to the velocity dispersions in the region of $20-100$ arcsec. \citet{Napolitano2014} only fit the upper limit of the velocity dispersion of the stellar kinematic data, which is $210-220$ \kms in that region. This is $\sim 10$ \kms higher that our data and our model prediction. This could explain the higher concentration of DM \citet{Napolitano2014} obtained. \citet{Napolitano2014} used similar GC data in the outer regions. The higher mass they obtained at larger radius could be a result of the mass-anisotropy degeneracy. As they obtained a higher mass as well as a more radially velocity dispersion anisotropy.   
   
Our estimated stellar mass-to-light ratio of $\Upsilon_V \sim$~8--9 may be compared with the
ATLAS$^{\rm 3D}$ result \citep{Cappellari2013b}. They found $\Upsilon_r = 7.0$, which we convert to $V$-band based on the galaxy's
$g-r = 0.9$ colour from SDSS, and arrive at $\Upsilon_V \simeq 8.5$, which is nicely consistent with our results. 
Spectroscopy-based stellar population synthesis modelling for NGC~5846 gives a stellar $\Upsilon_V =8.5$
assuming a Salpeter initial mass function (IMF; \citealt{Cappellari2013b}).
A Kroupa IMF would give $\Upsilon_V \simeq 5.3$, which we tentatively rule out with dynamic modelling, with the caveat that we have not allowed for adiabatically contracted halo models which would decrease the inferred stellar mass. With total stellar mass of $\sim 0.7\times 10^{12}\, M_{\odot}$, our galaxy is consistent with the trend of more massive galaxies displaying more Salpeter-like IMFs \citep{Cappellari2012}. Although things are actually more complicated, higher stellar mass-to-light ratio could also be accomplished by allowing the high-mass end slope of a Chabrier-like IMF to be steeper. As recently shown in \citet{Lyubenova2016}, the latter seems to be so far the preferred option for elliptical galaxies whereas a Salpeter-like single power-law is ruled out for $75\%$ of the galaxies in their sample. Moreover, there are also indications that the IMF might radially vary within galaxies \citep{Martn2015}. 
An interesting avenue for future work will be to combine our dynamical constraints with stellar population modelling that include IMF variations.
Our modelling result also agrees with the full-spectral fitting of the galaxy's central light with a free IMF \citep{Conroy2012}, which yielded $\Upsilon_V = 8.8$. 

The situation is thus different from the multi-population dynamical modelling of NGC~1407, where the difference between cored and cusped DM halo models 
was degenerate with the IMF assumption \citep{Pota2015b}.
For our NGC~5846 modelling, the primary limitation comes instead from the total dynamical mass estimate, rather than from the stellar mass-to-light ratio.

Overall, the dark and luminous mass distributions are individually well constrained in our model. 
Marginalizing over the different DM models, we find a stellar mass-to-light ratio of $\Upsilon_V = 8.8\pm0.5$.
We prefer a cored DM halo, which leads to only $\sim 10\%$ DM fraction within $1\,R_{\rm e}$, increasing to $67\pm10 \%$ at $6\,R_{\rm e}$. A standard NFW model is not favoured, although only with low statistical significance.

\section{Discussion: connections to other galaxies and formation clues}
\label{S:discussion}

\begin{figure}
\centering\includegraphics[width=\hsize]{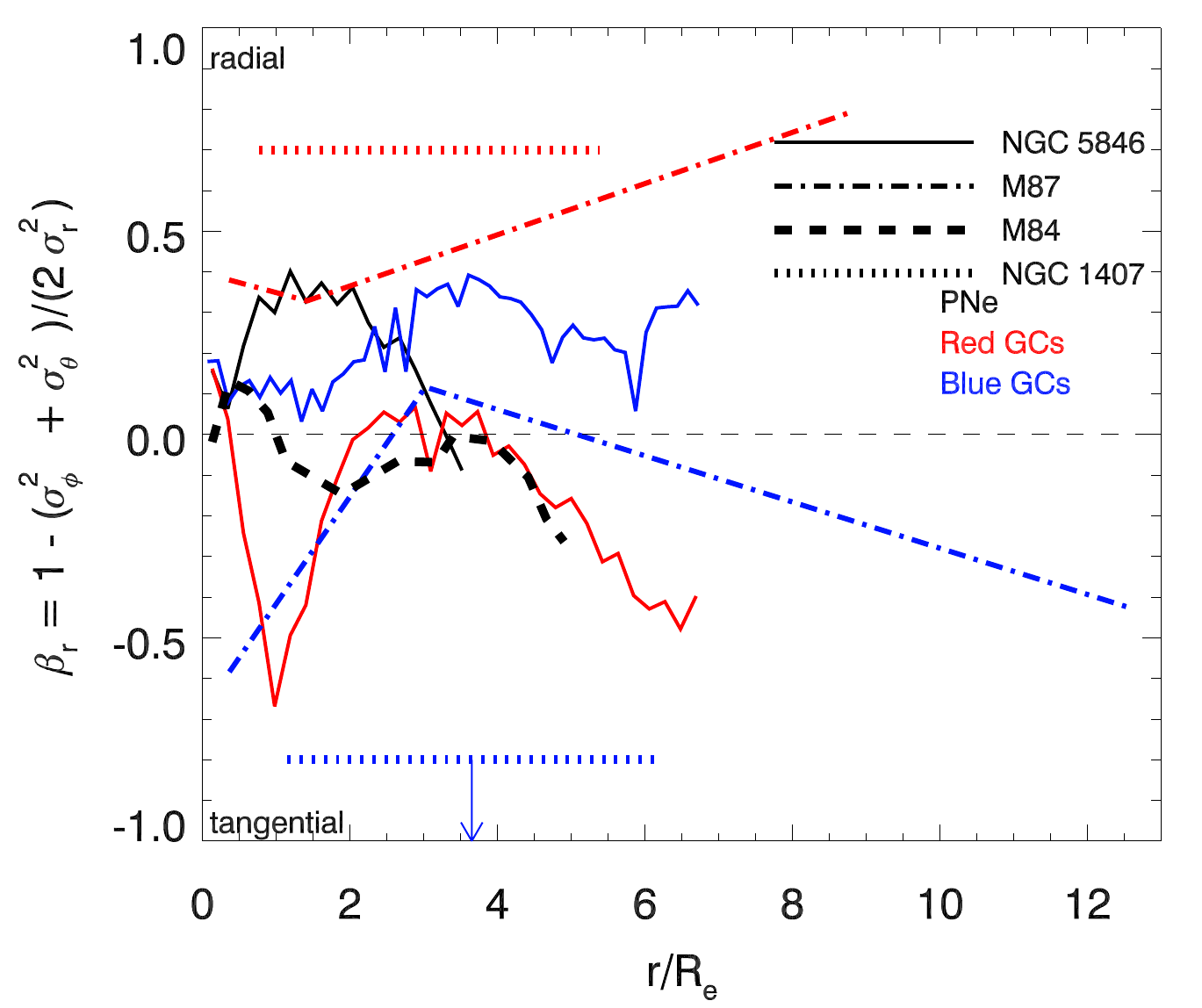}
\caption{The extended velocity anisotropy profiles of four giant ellipticals.  The solid lines represent the profiles for PNe, red GCs and blue GCs in NGC~5846 based on our M2M modelling, the dash-dotted lines represent red GCs and blue GCs in M87 from \citet{Zhang2015}, the dashed black line shows PNe in M84 from \citet{Zhu2014}, 
and the dotted horizontal lines are for red and blue GCs in NGC~1407 from \citet{Pota2015b}. }
\label{fig:betar_compare}
\end{figure}

We now consider some broader implications of our modelling results for NGC~5846.
In early-type galaxies, it is thought that the kinematics of red GCs are generally coupled to the field stars of the host galaxy, with rotation around the photometric minor axis, while blue GCs are more dominated by random motions \citep{Pota2013}. This is consistent with what we have found for NGC 5846, although we did not find the same red GC rotation 
as \citet{Pota2013} for this galaxy.
Considering the large fluctuations in rotation that we found  (Fig.~\ref{fig:xy_rot} and Fig.~\ref{fig:rot_fit}), 
the differences in results from \citet{Pota2013} could be due to their hard cut on the colour distribution in separating the two GC populations.  
In any case, both studies, as well as the PN study by \citet{Coccato2009}, agree on very low outer rotation in NGC~5846.
This result is consistent with cosmological simulations of massive galaxy formation \citep{Wu2014}, where slow rotation is a consequence of
multiple minor mergers from random directions (e.g. \citealt{Moody2014}).
A similar formation scenario for NGC~5846 is also inferred from a detailed analysis of its extended stellar kinematics by \citet{Forbes16}.

Additional clues to galaxy formation come from velocity dispersion anisotropy profiles, which we derived for NGC~5846 using 
three tracers: stars+PNe, red GCs, and blue GCs (Section~\ref{SS:anisotropy}).
These profiles can be compared to the generic expectations for elliptical galaxies that have formed through mergers. Such galaxies have mild anisotropy in their centres, yielding to strong radial anisotropy in their outer regions. This applies to both stars and GCs, whether they were formed in mergers or were passively accreted (e.g. \citealt{Dekel2005,Kruijssen2012}).
The blue GCs in NGC~5846 show this pattern, but the stars and red GCs, less so.
For the red GCs, the tangential anisotropy interior
to $\sim$~20~kpc  may simply relate to the effects of tidal forces, which preferentially dissolve the GCs that are on more radial orbits.
The isotropic to tangential trends for the stars and red GCs at $\gtrsim 30$~kpc
are more difficult to understand\footnote{\citet{Napolitano2014} inferred radial anisotropy for both red and blue GCs in NGC~5846 by spherical Jeans modelling using kurtosis constraints. However,  
by taking a spherical assumption, they lost the information of velocity dispersion anisotropy encoding the opening angle of the kinematic map (as we shown in Fig~\ref{fig:kin_map} and Fig~\ref{fig:kin}) and the uncertainties in the kurtosis were quite high.
In our discrete axisymmetric models, we have used the discrete data which encodes information about the opening angle of the kinematic map as well as the whole velocity distribution (for M2M; including the kurtosis). 
Therefore our anisotropy results should be more robust albeit still with substantial uncertainties. Considering the uncertainties of both sides, the \citet{Napolitano2014} results are still consistent with our results from M2M. },
but might connect to a component of {\it in situ} formation
\citep{Wu2014,Rottgers2014}.

In Figure~\ref{fig:betar_compare}, we compared results for the velocity anisotropy at large radii from NGC~5846 with other massive slow rotating ellipticals:
NGC~1407, M84, and M87.
For each galaxy, PNe, red GCs and blue GCs were treated as independent tracers. 
Among these four galaxies, there is a wide scatter of inferred anisotropy at both small and large radii,
with no consistent pattern within each galaxy or each tracer type.
The overall impression is of complex and inhomogeneous formation histories for this type of galaxy.
On the other hand, similar results for the field stars (including PNe) of fast-rotator ellipticals (NGC~3379, NGC~4494, NGC~4697, M60)
seem to have a more consistent pattern of isotropy in the centre, transitioning to radial anisotropy in the outer regions
as expected (based on field stars and PNe; \citealt{deLorenzi2008,deLorenzi2009,Das2011,Morganti2013}).
Tangential anisotropy for blue GCs from inner to outer regions is inferred from the negative kurtosis of the velocity distribution for the SLUGGS galaxies \citep{Pota2013}.

Given the uncertainties in our DM profile results for NGC~5846, considering their implications would take careful analysis that is beyond the scope of this paper.
However, we can make an interesting comparison of the {\it total} dynamical mass profile with the recent results of \citet{Cappellari2015} and \citet{Serra2016},
who studied a large sample of fast-rotator early-type galaxies using both extended stellar kinematics (from ATLAS$^{\rm 3D}$+SLUGGS) and HI kinematics.
They found a remarkable homogeneity in the total mass density profiles of these galaxies, with power-law slopes clustered tightly around $-2.2$ dex per dex.
The slope that we find for NGC~5846 over the radial range of 1--4~$R_{\rm e}$ is much shallower at $-1.7$~dex per dex.
This difference presumably arises from the location of NGC 5846 within a massive group-central halo, which is the dominant structure, but may also be a reflection of a difference in assembly histories between fast and slow rotators.

\section{Summary}
\label{S:conclusions}
We have applied our discrete chemo-dynamical modelling techniques to the giant elliptical galaxy NGC~5846, incorporating kinematic constraints from its red GCs, blue GCs and PNe -- along with field stars.
We used 214 GCs extending to $\sim 6 \,R_{\rm e}$ with LOS velocities from the SLUGGS survey, 123 PNe extending to $\sim 4\, R_{\rm e}$ with LOS velocities, SAURON IFU data extending to $\sim 0.5 \, R_{\rm e}$, and integrated stellar velocity and velocity dispersion measurements at 80 discrete positions extending to $\sim 1\,R_{\rm e}$ from SLUGGS.
In a departure from usual practice,
the GCs were {\it not} separated into red and blue GCs via a hard metallicity cut before modelling. Instead, a free colour distribution for each GC population was included in the 
chemo-dynamical model.
The three populations orbit in the same potential but with their own unique surface density profiles, internal rotation and velocity anisotropy properties.  By applying an MCMC process, the model was able to find, simultaneously, the region in parameter space that matches the observed kinematics. We have found:

\begin{itemize}
\item[(1)]  The mass profiles at all scales covered by the data points are constrained well. We constrain the total mass within $6\, R_{\rm e}$ ($=480'' = 57 \, \mathrm{kpc}$) to be $(1.7\pm0.3) \times 10^{12}\, M_{\odot}$, and obtain a $V$-band stellar mass-to-light ratio of $\Upsilon_V = 8.8\pm0.5$ (corresponding to a Salpeter IMF).
A cored DM halo is weakly preferred, and implies a DM fraction of $\sim 10\%$ within $1\,R_{\rm e}$, increasing to $67\pm 10 \%$ within $6\,R_{\rm e}$. 
A standard NFW halo, with $\sim 20\%$ DM within $1\,R_{\rm e}$, is not favoured although with only low statistical significance.

\item[(2)] The red and blue GCs are naturally separated out by the likelihood analysis.  We find weak rotation for the red GCs, with $v_{\rm max}/\sigma_0 \sim 0.3$ at $R> 1\, R_{\rm e}$. The rotation is about the photometric minor axis of the galaxy, and is opposite to the rotation of the inner field stars -- indicating a KDC.  We find no significant rotation for the PNe or blue GCs.

\item[(3)] We have created particle-based M2M models for the three populations, and used them to confirm that the velocity anisotropy profiles for the three populations as obtained by the axisymmetric Jeans models are acceptable, and are consistent with those obtained by the M2M models. 

The more general velocity anisotropy profiles obtained by the M2M models show that neither the red nor the blue GCs exactly follow the stars. 
In the inner regions, the red GCs are more tangentially anisotropic than the stellar tracers, which probably reflects
tidal forces destroying the GCs on radial orbits. 
In the outer regions, the blue GCs are radially anisotropic, as expected in accretion scenarios.
The outer PNe and red GCs show similar isotropic to tangential orbits. However, the red GCs have significant rotation which is not detected in the PNe or blue GCs.  
From their anisotropy and rotation, it raises the possibility that neither the red GCs nor the blue GCs have the same formation history as the stellar tracers. 
The outer PNe and red GCs show isotropic to tangential orbits, which is a feature seen in the PNe or blue GCs of other slow rotator ellipticals
\end{itemize}

\section{Acknowledements}
We thank XiangXiang Xue for useful discussions, and Alis Deason for sending her mass-profile results. 
Computer runs were mainly performed on the MPIA computer clusters \textit{queenbee} and \textit{theo}. This work was supported by Sonderforschungsbereich SFB 881 “The Milky Way System” (subprojects A7 \& A8) of the German Research Foundation (DFG) and the National Science Foundation grant AST-1211995, NRN is supported by Legge 5/2002 Regione Campania: "Dark Matter in Elliptical Galaxies", and DAF thanks the ARC for financial support via DP130100388.


\bibliographystyle{mn2e}
\bibliography{ngc5846}

\appendix
\section{Cusped dark matter models}
\label{S:kin_c5}

\begin{figure*}
\centering\includegraphics[width=\hsize]{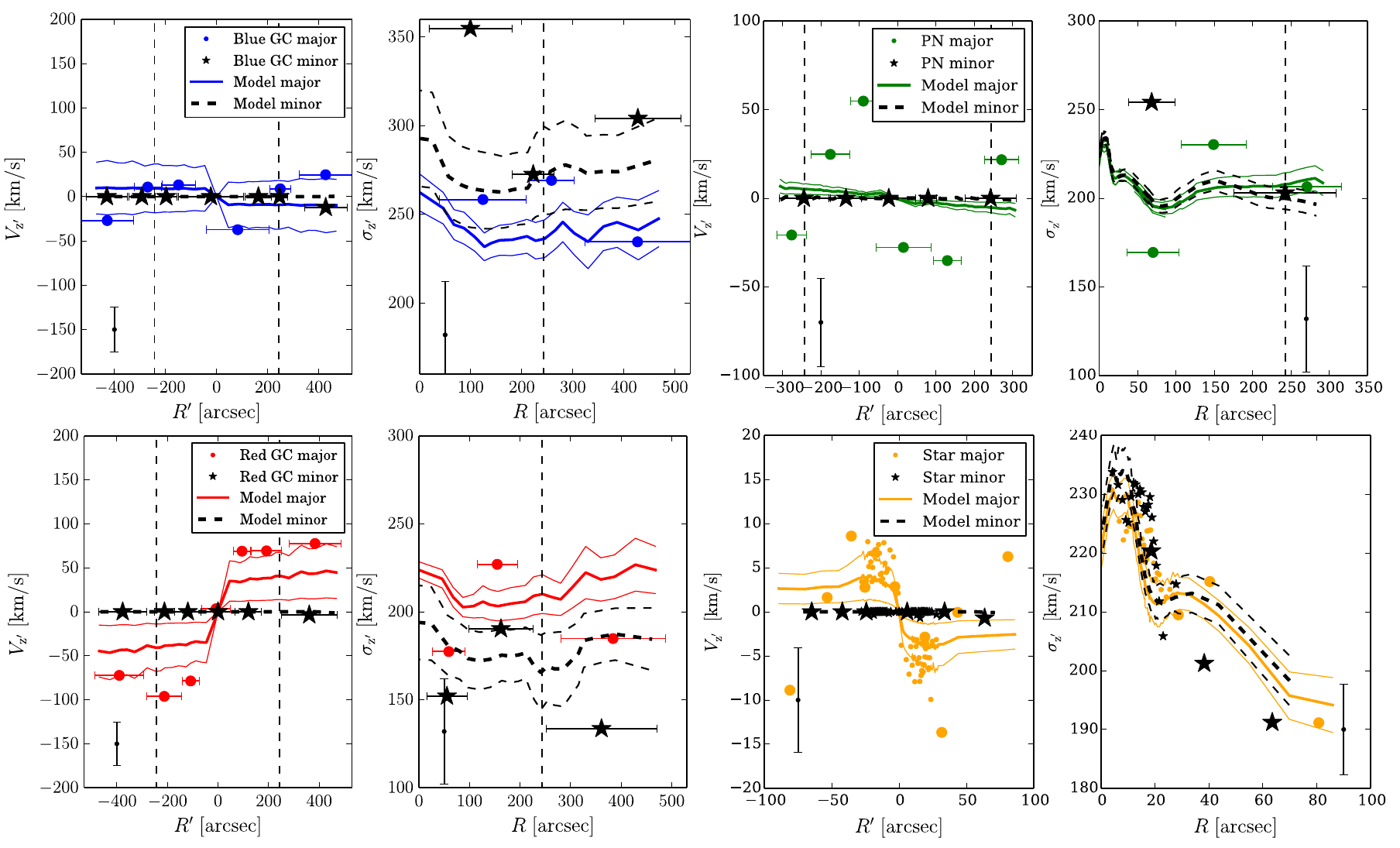}
\caption{Comparison between the data and the model predicted kinematics from the best-fitting cusped dark matter model, similar to Figure~\ref{fig:kin}. }
\label{fig:kin_cusp}
\end{figure*}

\begin{figure*}
\centering\includegraphics[width=\hsize]{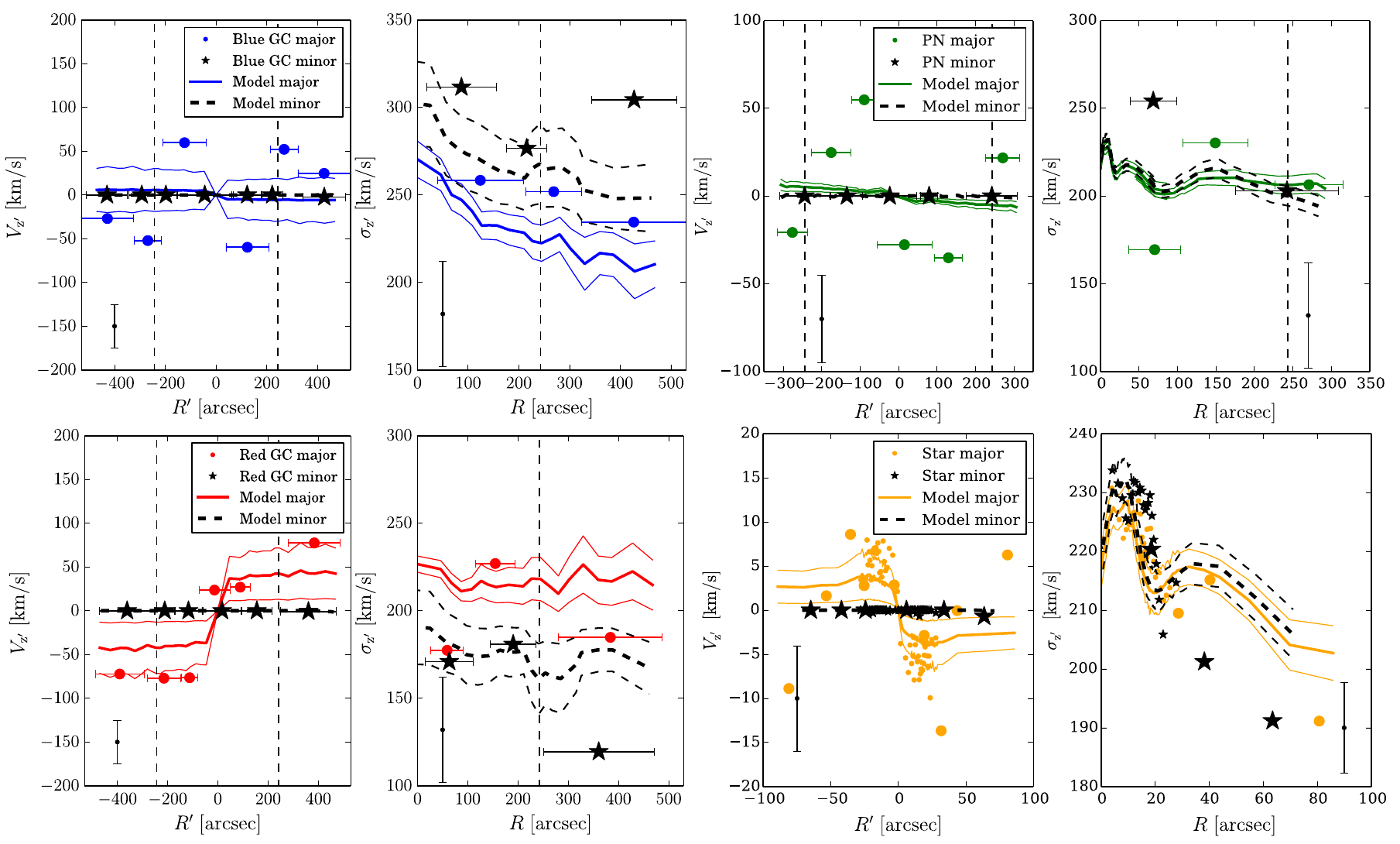}
\caption{Comparison between the data and the model predicted kinematics from the best-fitting constrained cusped model with $\rho_s > 1\times 10^{-3}\, M_{\odot}\,\mathrm{pc}^{-3}$. The model provides a poorer fit to the stellar kinematic data at $\sim 20-100$ arcsec. }
\label{fig:kin_c5}
\end{figure*}

Figure~\ref{fig:kin_cusp} and Figure~\ref{fig:kin_c5} show the fit to the kinematic data by the cusped model and the constrained cusped model with $\rho_s > 1\times 10^{-3}\, M_{\odot}\,\mathrm{pc}^{-3}$.  The data are binned in the same way as Figure~\ref{fig:kin}. The red and blue GCs are categorised slightly differently in different models, so that the kinematics of each population differ slightly between models.
Compared to the cored model in Figure~\ref{fig:kin}, the cusped model predicts a higher dispersion between $20-100$ arcsec, but still fits the data with almost equal quality.
The constrained cusped model predicts even higher dispersions between $20-100$ arcsec, and thus overfits the stellar kinematics  from the SLUGGS data in this region. At the same time, it predicts lower dispersions in the outermost regions, and thus the kinematics of the blue GCs are not fitted as well but the outer most data points of the red GCs are fitted somewhat better. 
With the large error bars in the data, we can see that, even for constrained cusped models, the model predictions are still consistent with the data within $1\sigma$ uncertainty. 
The differences in the velocity dispersions predicted by these three models are consistent with their differences in mass profiles. 

Note that the velocity dispersions from of SLUGGS data have already been adjusted by adding 10~\kms\ to match the SAURON data in the inner region. If we leave the SLUGGS data unchanged, then the constrained cusped model provides a poorer fit to the kinematics in this region. 

\section{The M2M models}
\label{S:m2m}
We create single-population M2M models for the red GCs, blue GCs and PNe respectively.  The models are created following \citet{Zhu2014}, to which we refer for further details. The same potential is used in all three models, and is the cored potential matching the black solid line in Fig~\ref{fig:mass}.  The M2M models are not created to determine any potential parameters, but to investigate the dynamical properties of each tracer population.
The surface number density and discrete LOS velocities of the corresponding tracers are used as constraints in each model. 

\subsection{The initial conditions}
We initially give the particles equal luminosity weights. The particle number density profile follows the 3D number density profile deprojected from the observed surface number density for each population. The velocity distribution at each radius is a Gaussian distribution, with a mean velocity of zero and velocity dispersion following the solution of an isotropic, spherical Jeans model using the M2M potential and the population tracer number density profile. No net rotation of the particle system is used.  In total, we create 400000 particles for each model. 

\subsection{Fit to the data}
Discrete LOS velocity data from our GCs and PNe are included in the model. Overall we have 100 red GCs, 96 blue GCs, and 123 PNe. To reduce noise from the data, axis-symmetrization and point-symmetrization (e.g., \citealt{deLorenzi2008}; \citealt{vdB2008}) are applied with the data sizes being enlarged by a factor of 4 as a result. 

\begin{figure}
\centering\includegraphics[width=6cm]{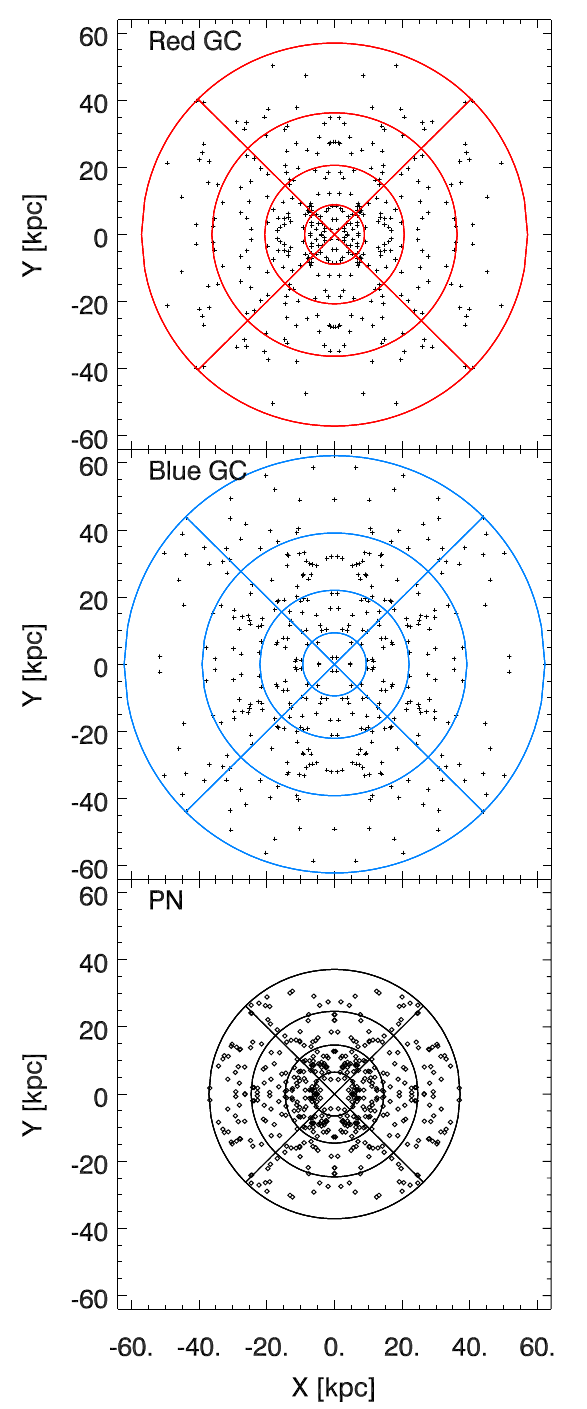}
\caption{The binning schemes for the M2M models of the red GCs, blue GCs and PNe from top to bottom. 
The black points in each panel represent the positions of the discrete data, the circles indicate the binning in radius and the solid lines represent the binning in azimuthal angle for each model. The binning schemes about radially are different from each other in the three models.}
\label{fig:bin_scheme}
\end{figure}

Model data values are formed by binning the particle data onto the projected plane using $4\time4$ bins in projected radius and azimuthal angle as shown in Fig~\ref{fig:bin_scheme}.  The binning schemes are constructed in the same way for each model, differing only in bin radius because the three tracers cover slightly different spatial regions.

Given the small number of data points available, regularisation is used to ensure indirectly a smooth final velocity distribution \citep{Long2010}.  Models with different regularisation were tried, and we show here three models with high regularisation ($\mu = 1.5$), low regularisation ($\mu = 0.5$) and with no regularisation ($\mu = 0.0$). The absolute value of the parameter $\mu$ has no physical meaning: it merely sets the relative influence of regularisation with reference to the other modelling constraints.  Parameter $\mu$'s role is not dissimilar to parameter $\alpha_s$'s role in Section~\ref{SS:combine}.

\begin{figure}
\centering\includegraphics[width=10cm,height=6cm]{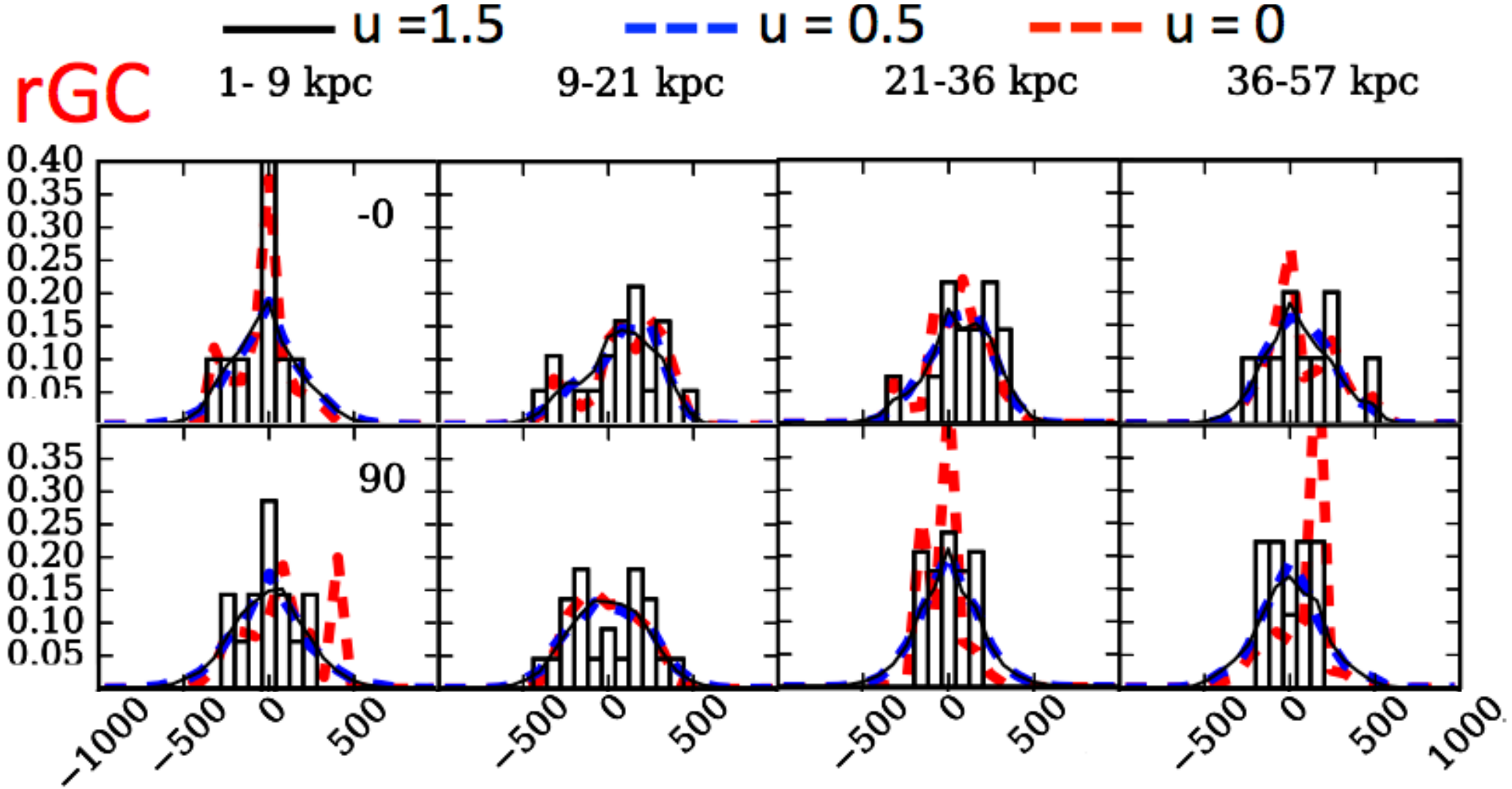}
\includegraphics[width=10cm,height=5.5cm]{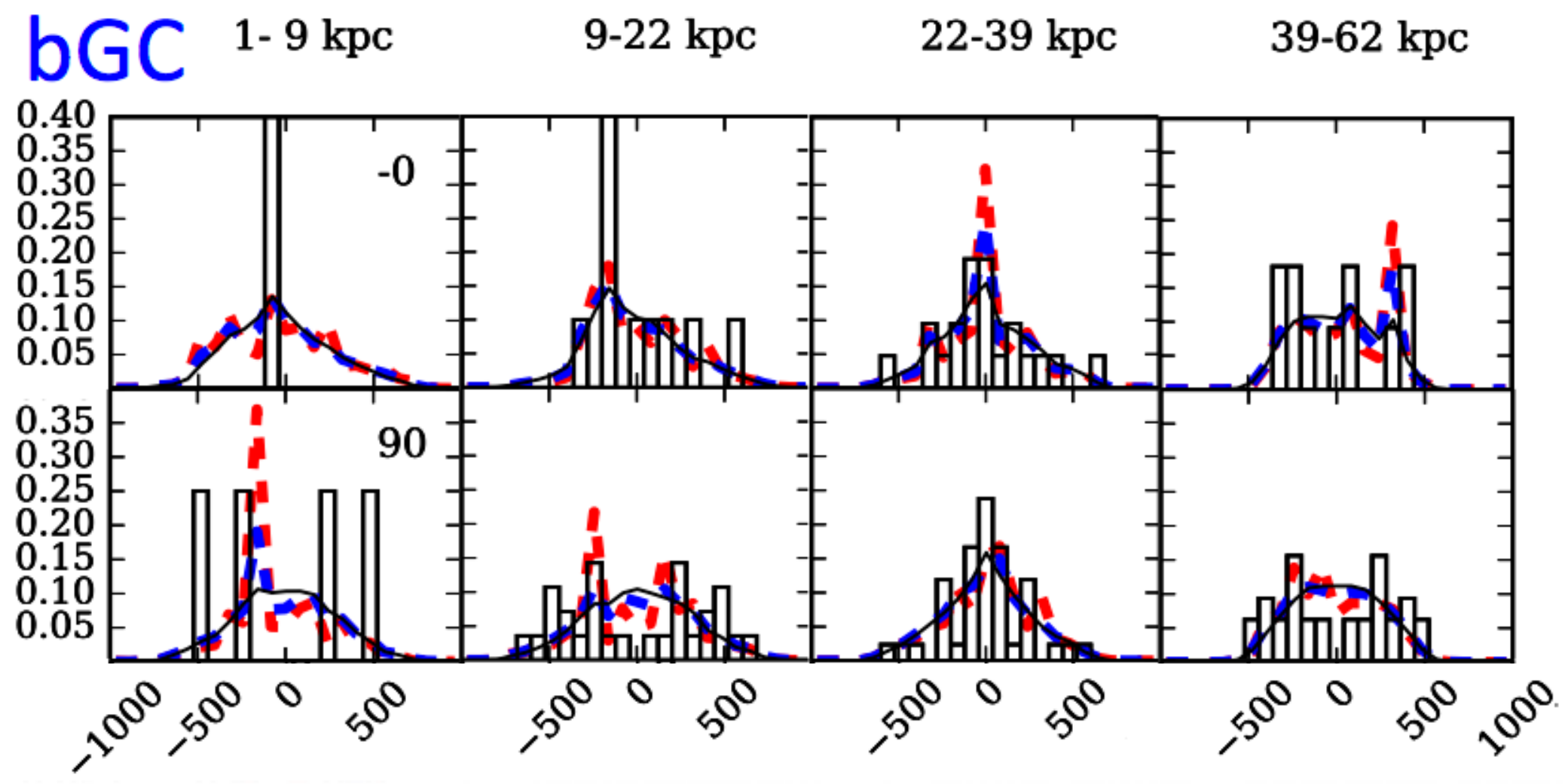}
\includegraphics[width=10cm, height=5.5cm]{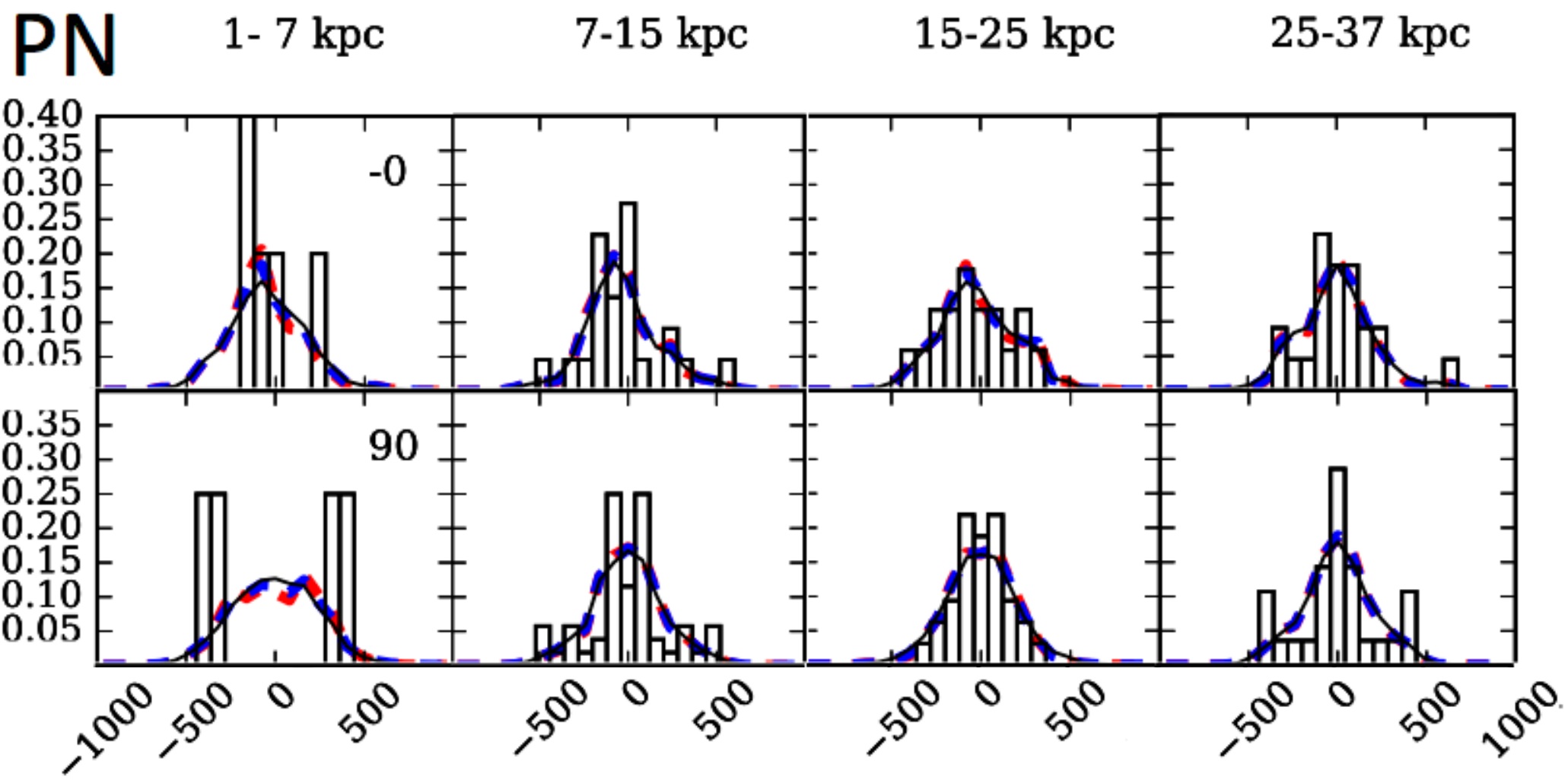}
\caption{The velocity distributions for the models of red GCs, blue GCs and PNe from top to bottom. The histograms are the velocity distributions of the data in each bin. The black solid lines, blue dashed lines and red dashed lines are those of the models with regularisation $\mu = 1.5, \, 0.5, \, 0.0$ respectively. 
}
\label{fig:M2Mfit}
\end{figure}

For each model, the particle weights converged well with more than $95\%$ of particles having weight variations less than $5\%$ in the last 20 half mass dynamical time units of modelling.
Fig~\ref{fig:M2Mfit} shows how well the M2M models fit the data for the models of the red GCs, blue GCs and PNe reading from top to bottom. We only show the velocity distributions for the bins with a central azimuthal angle of $0^{\circ}$ or $90^{\circ}$. The other bins are not shown, as the velocity distributions are similar but with oppositely signed velocities in the bins with an azimuthal angle difference of $180^{\circ}$.
The models with no or little regularisation are quite noisy and have some local peaks in the velocity distributions. The model with $\mu = 1.5$ is generally smooth.  The velocity anisotropy profiles we use in Fig~\ref{fig:beta} are directly calculated from the highly regularised models.

\bsp 

\label{lastpage}

\end{document}